\documentclass[final,3p,times,11pt]{elsarticle}
\journal{Physica D}

\usepackage{mathcomp} 
\usepackage{textcomp}
\usepackage{graphicx}
\usepackage{amsmath}
\usepackage{graphicx}
\usepackage{amssymb}
\usepackage{xcolor}
\usepackage{showlabels} 

\newcommand{\lbl}[1]{\label{#1}} 

\newfont{\bbold}{msbm10 scaled\magstep1}
\newcommand{\qtilde}[1]{}

\newcommand{\etal}{{\em et al.}}

\newcommand{\nn}{\nonumber}
\newcommand{\dd}{{\rm d}}

\newcommand{\ep}{\epsilon}

\newcommand{\rec}[1]{\mbox{$\frac{1}{#1}$}}
\newcommand{\mfrac}[2]{\mbox{$\frac{#1}{#2}$}}
\newcommand{\half}{\rec{2}} 
\newcommand{\ee}{{\rm e}}
\newcommand{\eee}[1]{\ee^{{#1} (i k m + i l h n - i \omega t)}}
\newcommand{\beq}{\begin{equation}} 
\newcommand{\eeq}{\end{equation}} 
\newcommand{\beqa}{\begin{eqnarray}}
\newcommand{\eeqa}{\end{eqnarray}}

\newcommand{\sech}{{\rm sech}}
\newcommand{\vect}[3]{\begin{pmatrix} #1 \\ #2 \\ #3 \end{pmatrix}}
\newcommand{\mat}[9]{\begin{pmatrix} #1 & #2 & #3 \\ #4 & #5 & #6 \\  #7 & #8 & #9 \end{pmatrix}}

\newcommand{\qomit}[1]{}
\newcommand{\bfk}{\mbox{${\bf k}$}}

\newcommand{\blue}[1]{{#1}}
 \newcommand{\purple}[1]{{#1}}

\begin{document}

\title{{\bf Asymptotic Analysis of discrete nonlinear localised modes 
in a kagome lattice. }} 

\author{Jonathan AD Wattis$^{*1}$,  
Pilar R Gordoa$^2$, 
Andrew Pickering$^3$
\\
{\footnotesize 
$^1$School of Mathematical Sciences, University of Nottingham, 
University Park, Nottingham NG7 2RD, 
UK }\\[-0.4ex]
{\footnotesize $^{2,3}$\'{A}rea de Matem\'{a}tica Aplicada, ESCET, Universidad 
Rey Juan Carlos, C/Tulip\'{a}n s/n, 28933 M\'{o}stoles, Madrid, Spain}\\[-0.4ex] 
{\footnotesize 	$^1$ Jonathan.Wattis@nottingham.ac.uk \qquad 
* Corresponding Author}\\[-0.4ex]
{\footnotesize $^2$ Pilar.Gordoa@urjc.es \qquad 
$^3$ Andrew.Pickering@urjc.es } }

\date{\footnotesize{\today}}

\begin{abstract}
We describe a nonlinear kagome lattice with nonlinear 
dynamics described by Klein-Gordon interactions with a 
scalar unknown at each node, such as might occur in a 
nonlinear electrical lattice.  We show that the dispersion relation 
has three bands - a flat band and two other surfaces which 
may meet in Dirac points or be separated by a gap. 
By using multiple scales asymptotic methods, we find a 
variety of reductions to nonlinear Schr\"{o}dinger (NLS) systems, 
some of which are similar to those obtained previously, 
and have the Townes soliton as a solution. 
\blue{We find a novel system of coupled NLS equations, 
by forming an asymptotic expansion for small amplitude 
weakly nonlinear waves around the point where the flat 
band meets the upper surface of the dispersion relation.  
We analyse this 2+1 dimensional system} 
using Lie symmetries, and find further reductions to 
more complicated solitary wave solutions.  
\blue{Numerical simulations of the wave are also presented.} 
\end{abstract}

\maketitle

\subsubsection*{{\bf Highlights}} 

\noindent$\bullet$ We derive the nonlinear Klein-Gordon kagome lattice equations. \\ 
\noindent$\bullet$ We use asymptotic techniques to find reductions to nonlinear Schr\"{o}dinger systems. \\  
\noindent$\bullet$ We use Lie symmetries to find reductions 
of PDEs to ODEs. \\[2ex]  
\noindent{\bf Keywords} \\
Kagome lattice, coupled NLS, discrete breathers, nonlinear localised modes, asymptotic analysis.  

\section{Introduction} 
\lbl{intro-sec}

\blue{We analyse small amplitude weakly nonlinear oscillations in 
a kagome lattice with Klein-Gordon interactions, that is, 
linear nearest-neighbour interactions and a nonlinear on-site 
potential.  We propose a simple model system that has 
the kagome lattice structure, we only consider a single 
scalar unknown quantity at each node. Using multiple scales 
asymptotic techniques, 
we seek breather solutions, which have the form of an 
envelope for a linear carrier wave.  We reduce the system to 
nonlinear Schr\"{o}dinger (NLS) form; for some choices of 
wavenumbers we obtain novel coupled system of NLS equations 
in two space dimensions, for which we obtain similarity 
solutions using Lie point symmetries. }

Kagome lattices exhibit star-shaped hexagonal symmetry, their 
lattice being composed of hexagons and triangles (there being 
twice as many of the latter), as shown in Figure \ref{k-latt-fig}. 
Amongst two-dimensional lattices, after square lattices, 
the most commonly studied are those with hexagonal symmetry; 
the triangular lattice and honeycomb are the simplest, 
but in recent years, there has been significant interest in the 
properties of the kagome lattice.  
Many lattice systems have dispersion relations with multiple 
surfaces; this is due to there being multiple degrees 
of freedom within each unit cell (either multiple nodes in 
a cell, or nodes being displaced in multiple directions, 
or both).
The simplest kagome lattices have three surfaces, since there 
are 3 nodes in the unit lattice cell. 
For certain choices of parameter values, 
two surfaces meet in the form of a double cone, 
which is described as a `Dirac point'. 
In addition to energy propagation through nonlinear  lattices, 
the kagome  model is of interest due to it being less regular 
than the triangular or honeycomb model, so that it may display 
some similarities with more random networks;  it has been 
applied to the study of order/disorder transitions in voting models 
\cite{mvm} and in the Potts model \cite{Potts}. 

\subsection{Mathematical Results}

The rigorous proof of the existence of breathers in discrete lattices 
was originally formulated by MacKay and Aubry \cite{MA}, for 
one-dimensional systems, using the anticontinuum limit, a method 
which generalises to a wide range of higher dimensional systems. 
Flach and Gorbach \cite{FG} give a review of discrete breathers 
which covers higher dimensional lattices as well as 1D chains. 

Motivated by the work of Marin \etal\ \cite{jce-sq,jce-hex} 
who performed some of the first numerical simulations which 
illustrated breathers moving through 2D lattices, we have 
previously use asymptotic expansions to study the nonlinear 
Klein-Gordon and FPUT 
equations on square \cite{alz}, triangular \cite{tri} and honeycomb 
\cite{honey,mica} lattices. We have considered both cases where 
there is a single unknown at each node (as occurs, for example, 
in electrical transmission lattices, \blue{where the transport of charge 
is of primary interest}), and mechanical systems, where 
there are two unknown displacements at each node, and where 
displacements in the two directions within the lattice are coupled, 
leading to more complicated governing equations of motion. 

More recent theoretical results includes the work of Hofstrand \etal\ 
\cite{HW} who consider the simpler dimer form of the FPUT lattice 
and use multiple scales asymptotic techniques to construct 
an approximate solution in the gap in dispersion relation. 
Hofstrand \cite{Hof} considers a kagome lattice using 
numerical techniques to find breathers at certain wave-numbers, 
and shows their stability via the calculation of Floquet multipliers.  
\blue{In further work, Hofstrand \cite{hof26} analyses a 
modified hexagonal lattice in which nearest-neighbours in 
one direction (e.g.~vertical) have different interactions to the other 
two (diagonal) directions. He obtains a dispersion relation 
with a `semi-Dirac' point, that is, it has two surfaces, 
which cross linearly in one direction, but quadratically 
in the other direction.   He shows that stable 
breather modes can exist in this system, and investigates 
the bifurcations through which they lose stability. } 
Below, we consider the a similar system, and use asymptotic 
techniques to show how breathers are related to solutions 
of the 2D nonlinear Schr\"{o}dinger (NLS) system. 
\blue{Shi \etal\ \cite{shi} present results on the stability of 
compactly-supported solitons on lattice systems with a 
flat band and quote the DNLS equation on a kagome 
lattice as a particular example. }

Babicheva \etal\ \cite{bab} consider discrete breathers on 
a FPUT lattice  with triangular symmetry.  They seek nonlinear 
modes with frequencies above the maximum of the phonon band. 
As well as fully localised breathers, they also find various modes 
which are localised in one direction, but delocalised in the other, 
these they term delocalised nonlinear vibrational modes (DNVM). 
They form ad hoc approximations of these modes.  
\blue{Kirsch \etal\ \cite{kirsch} demonstrate the 
existence of nonlinear corner-state modes in 
photonic insulators with a kagome lattice structure, 
which is finite and triangular, meaning the corner is acute (60$^\circ$). 
They model the two-dimensional topological insulator  
using a 2+1 dimensional NLS equation with an additional 
linear potential term which describes the lattice. 
They explore a range of nearest-neighour coupling strengths 
(similar to varying the magnitudes of $\lambda$, $\gamma$ 
in our model below). 
They simulate the system numerically, as well as presenting images 
from experiments, which show good agreement across a range of 
powers and amplitudes which explore both weak and more strongly 
nonlinear regimes.}  

Vicencio and Johansson \cite{VJ} consider a coupled lattice 
of nonlinear Schr\"{o}dinger equations which has a similar 
dispersion relation to that considered here. They perform a 
bifurcation analysis of the system and find 
families of localised discrete solitons bifurcating from 
the flat band in the case of defocusing nonlinearity.  

\subsection{Applications of density functional theory to kagome lattices}

\blue{Molecular structure at the microscale affects physical properties 
in many ways, some of which are complicated and subtle, and so 
detailed approaches using density functional theory (DFT) has 
been used to investigate such effects.  }
\blue{Uzunok \etal\ \cite{uzunok} report the results of {\em ab initio} 
density functional theory calculations of a range of 
superconductors with a hexagonal lattice structure that 
includes a layer of transition metal atoms with kagome geometry.  
They find that superconductivity stems from the propagation 
of phonons and of electron-phonon interactions in this layer. }
\blue{Im \etal\ \cite{im} employ density functional theory to investigate 
the properties of a monolayer kagome lattice (in particular, 
AV$_3$Sb$_5$, with A=K,Rb,Cs).   They find that charge-density 
waves exist in a monolayer, and that such waves are more stable 
in a monolayer than in bulk. 
This extends the earlier work of Ferrari \etal\ \cite{ferrari}, who 
used the Hubbard model to investigate charge-density waves in 
AV$_3$Sb$_5$ (A=K,Rb,Cs) materials.  Their numerical results show 
an instability with the electron-phonon coupling causing a distortion 
to the lattice, in agreement with experimental findings. 
Fu \etal\ \cite{fu} investigate the role of loop currents and 
competitive charge instabilities in a kagome lattice. }
\blue{Liu \etal\ \cite{liu} consider a fundamental model lattice structure, 
with kagome geometry.  They find minor distortions result in changes 
to the dispersion relation, the density of states, and optical conductivity. 
They calculate second-order nonlinear optical response 
for the example crystal lattice of Mo$_5$Te$_8$. }
Singh \& Garcia-Castro \cite{Singh} also use DFT  
to investigate the propagation of charge-density waves through 
kagome lattices to investigate theoretically the remarkable 
material properties observed in metal alloys with this structure. 
Prabith \etal\ \cite{prabith}  analyse the behaviour of 
nodes near the vertex of a finite kagome lattice with overall 
triangular shape.  They find a variety of localised modes 
(with 2 or 3 excited states), and explore these using numerical 
path-following bifurcation packages,  finding both stable and 
unstable modes. 

\subsection{Other applications}

The properties of kagome lattices can also be exploited in larger-scale 
structures.   Motivated by material properties, 
Law \etal\ \cite{LAW} investigate the existence and stability 
of gap vortices in a kagome lattice with external modulation, 
their governing equations having the form of a nonlinear 
Schr\"{o}dinger system with defocusing nonlinearity. After considering 
conditions for the existence and stability of modes, they show 
numerical simulations of a variety of localised structures 
(in-phase and out-of-phase, vortices, quadrupole, and hexapoles), 
some of which are stable and others unstable. 
\blue{Augello \etal\ \cite{augello} consider the mechanical behaviour 
of a metamaterial whose composition is based on the kagome lattice. 
They use finite element methods to compute the stress-strain 
relationships of a one-dimensional beam with a kagome microstructure, 
and find failure modes which rely on hinging about certain parts 
of the lattice, and hence make suggestions on which hinge-points of the 
structure need strengthening in order to improve the large-scale integrity 
of the beam. }
\blue{Yang \& Wang \cite{yw} investigate the elastic properties 
of metamaterials with internal kagome microstructure 
composed of a hexagonal array of triangles that can rotate within 
the lattice. This leads to a highly complicated dispersion relation 
and elastic waves with unusual properties due to multiple 
degeneracies in the mass-spring model. } 
Chern \etal\ \cite{chern} investigate an optical lattice system 
modelled by a quantum Hamiltonian, and compare the 
flux through square and kagome lattices. 
Lee \cite{lee} considers a linear Schr\"{o}dinger equation 
with point scatterers arranged in a honeycomb lattice. 
The presence of Dirac points in the dispersion relation 
is discussed in the context of the electronic 
properties of graphene, and numerical simulations are 
presented for a range of parameter values. 
For a more detailed review of the physical properties associated 
with kagome lattices, we refer the reader to Di Sante \etal\ \cite{dis}. 

\subsection{Outline} 

In section \ref{asy-sec} we outline the precise model studied 
herein -- a kagome lattice with an unknown scalar at each node.
We outline the asymptotic reduction from the Klein-Gordon  
model to NLS equations in Section \ref{asy-sec}. 
\blue{This method is based on the assumption of small amplitude 
oscillations, since it makes use of methods which are 
valid in the weakly nonlinear limit.}
One particular case leads to a novel 
coupled system of NLS equations, which we study in more detail in 
Section \ref{simil-sec}, where similarity reductions are obtained.  
Conclusions are discussed in the final section. 

\section{Asymptotic analysis of nonlinear kagome lattice} 
\lbl{asy-sec}
\setcounter{equation}{0}

\subsection{Formulation of Klein-Gordon system}

\begin{figure}[ht] 
\begin{picture}(400,170)(-30,-20)
\thicklines
\put(-10,  0){\line(1,0){290}}
\put(-10, 50){\line(1,0){290}}
\put(-10,100){\line(1,0){290}}
\put(-10,150){\line(1,0){290}}
\put( -3,95){\line(3,5){40}}
\put( -3,-5){\line(3,5){100}}
\put( 57,-5){\line(3,5){100}}
\put(117,-5){\line(3,5){100}}
\put(177,-5){\line(3,5){100}}
\put(237,-5){\line(3,5){40}}
\put( 33,-5){\line(-3,5){40}}
\put( 93,-5){\line(-3,5){100}}
\put(153,-5){\line(-3,5){100}}
\put(213,-5){\line(-3,5){100}}
\put(273,-5){\line(-3,5){100}}
\put(273,95){\line(-3,5){40}}
\thinlines 
\multiput(0,-10)(5,0){60}{\circle*{1}}
\multiput(0, 40)(5,0){60}{\circle*{1}}
\multiput(0, 90)(5,0){60}{\circle*{1}}
\multiput(0,140)(5,0){60}{\circle*{1}}
\multiput(15,-13)(60,0){4}{\circle*{1}}
\multiput( 15,40)(0,5){10}{\circle*{1}}
\multiput( 75,40)(0,5){10}{\circle*{1}}
\multiput(135,40)(0,5){10}{\circle*{1}}
\multiput(195,40)(0,5){10}{\circle*{1}}
\multiput(255,40)(0,5){10}{\circle*{1}} 
\multiput( 15,140)(0,5){4}{\circle*{1}}
\multiput( 75,140)(0,5){4}{\circle*{1}}
\multiput(135,140)(0,5){4}{\circle*{1}}
\multiput(195,140)(0,5){4}{\circle*{1}}
\multiput(255,140)(0,5){4}{\circle*{1}} 
\multiput( 45,90)(0,5){10}{\circle*{1}}
\multiput(105,90)(0,5){10}{\circle*{1}}
\multiput(165,90)(0,5){10}{\circle*{1}}
\multiput(225,90)(0,5){10}{\circle*{1}}
\multiput( 45,-10)(0,5){10}{\circle*{1}}
\multiput(105,-10)(0,5){10}{\circle*{1}}
\multiput(165,-10)(0,5){10}{\circle*{1}}
\multiput(225,-10)(0,5){10}{\circle*{1}}
\multiput(11,21)(60,0){5}{$\triangledown$}
\multiput(41,71)(60,0){4}{$\triangledown$}
\multiput(11,121)(60,0){5}{$\triangledown$}
\multiput(30,  0)(60,0){5}{\circle{5}}
\multiput( 0, 50)(60,0){5}{\circle{5}}
\multiput(30,100)(60,0){5}{\circle{5}}
\multiput(0,150)(60,0){5}{\circle{5}}
\multiput(0,0)(60,0){5}{\circle*{5}}
\multiput(0,100)(60,0){5}{\circle*{5}}
\multiput(30,50)(60,0){5}{\circle*{5}}
\multiput(30,150)(60,0){5}{\circle*{5}}
\put( 90,62){$q^{*}_{m,n}$}
\put( 25,62){$q^{*}_{m-2,n}$}
\put(145,62){$q^{*}_{m+2,n}$}
\put(55,9){$q^{*}_{m-1,n-1}$}
\put(55,113){$q^{*}_{m-1,n+1}$}
\put(115,9){$q^{*}_{m+1,n-1}$}
\put(115,113){$q^{*}_{m+1,n+1}$}
\put(110,69){$A$}
\put( 77,52){$B$}
\put(117,54){$C$}
\put(137,51){$B$}
\put(59,54){$C$}
\put(117,25){$A$}
\put(80,25){$A$}
\put(125,91){$B$}
\put(76,91){$C$}
\end{picture}
\caption{Illustration of kagome lattice - triangles indicate A-nodes, 
filled circles B-nodes, and open circles, C-nodes. 
Edges of unit cells are denoted by dotted lines.  
\lbl{k-latt-fig} } 
\end{figure}
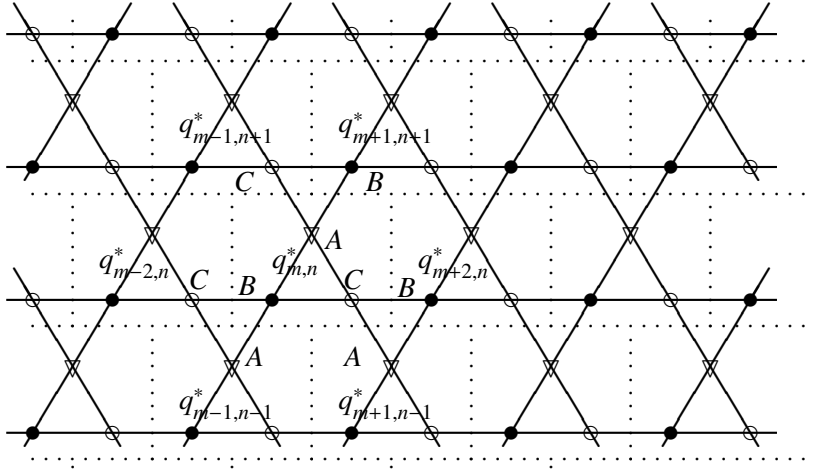 

The kagome lattice is illustrated in Figure \ref{k-latt-fig}. 
Each cell contains three nodes, one of each type, $A,B,C$.  
Cells have six neighbouring cells, but each node is coupled to only four 
neighbouring nodes, two in the same cell and two in two different 
neighbouring cells. Each node is coupled to nodes of the other type, 
e.g.~$q_{m,n}^B$ is coupled to two $A$-nodes and two $C$-nodes, 
and no other $B$-node. 
All node-node interactions are assumed to be linear; 
the coupling within a cell has strength $\gamma$, and between cells $\lambda$, 
and in general we assume $\lambda\neq\gamma$, but the case $\gamma=\lambda$ 
is a special case that we will also analyse. In addition to the nearest neighbour 
interactions, we assume a nonlinear on-site potential given by quartic anharmonicity 
of strength $\beta$, with harmonic term $\Omega^2$ giving the Hamiltonian 
\begin{align}
H = & \; \sum_{ \genfrac{}{}{0pt}{1}{(m,n)}{m+n\;\mbox{\scriptsize even}} } 
\left\{ 
\frac12 \left( \!\frac{\dd q^A_{m,n}}{\dd t} \right)^2 + 
\frac12 \left( \!\frac{\dd q^B_{m,n}}{\dd t} \right)^2  + 
\frac12 \left( \!\frac{\dd q^C_{m,n}}{\dd t} \right)^2  
+ \frac{ \Omega^2}{2} \left( (q_{m,n}^{A})^2+(q_{m,n}^{B})^2+(q_{m,n}^{C})^2 
\right)  \right. \nn \\ & \left.  \qquad \qquad 
+ \frac{\beta}{4} \left( (q_{m,n}^{A})^4+(q_{m,n}^{B})^4+ (q_{m,n}^{C})^4 \right) 
+ \frac{\gamma}{2} \left[  (q_{m,n}^A - q_{m,n}^B )^2 
+ ( q_{m,n}^B - q_{m,n}^C )^2 
+ ( q_{m,n}^C - q_{m,n}^A )^2 \right] 
\right. \nn \\ & \left.  \qquad \qquad 
+ \frac{\lambda}{2} ( q_{m,n}^A - q_{m+1,n+1}^B )^2
+ \frac{\lambda}{2} ( q_{m,n}^A - q_{m-1,n+1}^C )^2
+ \frac{\lambda}{2} ( q_{m,n}^B - q_{m-2,n}^C )^2
\genfrac{}{}{0pt}{0}{ \ }{ \ } \right\} . \lbl{ham}
\end{align}
The equations of motion are thus 
\begin{align}
\frac{\dd^2 q_{m,n}^A}{\dd t^2} = & - \Omega^2 q_{m,n}^A - \beta (q_{m,n}^A)^3 
+ \gamma ( q_{m,n}^B + q_{m,n}^C - 2 q_{m,n}^A ) 
+ \lambda ( q_{m+1,n+1}^B + q_{m-1,n+1}^C - 2 q_{m,n}^A ) , 
\lbl{qAddot} \\[0.5ex]
\frac{\dd^2 q_{m,n}^B}{\dd t^2} = & - \Omega^2 q_{m,n}^B - \beta (q_{m,n}^B)^3 
+ \gamma ( q_{m,n}^C + q_{m,n}^A - 2 q_{m,n}^B ) 
+ \lambda ( q_{m-2,n}^C + q_{m-1,n-1}^A - 2 q_{m,n}^B ) ,  \\[0.5ex]
\frac{\dd^2 q_{m,n}^C}{\dd t^2} = & - \Omega^2 q_{m,n}^C - \beta (q_{m,n}^C)^3 
+ \gamma ( q_{m,n}^A + q_{m,n}^B - 2 q_{m,n}^C ) 
+ \lambda ( q_{m+1,n-1}^A + q_{m+2,n}^B - 2 q_{m,n}^C ) . \lbl{qCddot} 
\end{align}

\subsection{\blue{Dispersion relation}} 

\blue{We start by looking at the dispersion relation for small 
amplitude waves. 
We write $q_{m,n}^K(t) = \ep F_K \eee{}+c.c.$, with constants 
$F_K$ ($K=\{A,B,C\}$) denoting relative amplitudes, $c.c.$ 
indicating complex conjugate of preceding terms,  and 
$\ep\ll1$ indicating that all three quantities undergo small 
amplitude oscillations.  This has the effect of removing the 
nonlinear term from the system (\ref{qAddot})--(\ref{qCddot}), 
equivalent to setting $\beta=0$. The wavenumbers in the $x$- and 
$y$-directions are denoted by real quantities $k,l$, and the 
temporal frequency is given by $\omega=\omega(k,l)$.  
For notational simplicity, we introduce $h=\sqrt{3}$ 
so that the distance between neighbouring nodes 
is the same in the horizontal direction as along the two diagonals 
(that is, the distance between nodes $(m,n)$ and $(m\pm2,n)$ 
is that same as that between $(m,n)$ and $(m\pm1,n\pm1)$). }

From (\ref{qAddot})--(\ref{qCddot}) we obtain the system 
of coupled equations for $F_A,F_B,F_C$ 
\begin{align}
{\bf M f} := 
\mat{\omega^2\!-\!2\lambda\!-\!2\gamma \!-\!\Omega^2  }
{\gamma + \lambda \ee^{ik+ilh} }{\gamma + \lambda \ee^{-ik+ilh} }
{\gamma + \lambda \ee^{-ik-ilh}}{\omega^2\!-\!2\lambda\!-\!2\gamma 
\!-\!\Omega^2 }{\gamma + \lambda \ee^{-2ik}}
{\gamma + \lambda \ee^{ik-ilh}}{\gamma + \lambda \ee^{2ik}}
{\omega^2\!-\!2\lambda\!-\!2\gamma \!-\!\Omega^2 } 
\!\vect{F_A}{F_B}{F_C} \!=\! \vect{0}{0}{0}\!.  
\lbl{mf-equal-zero} \end{align}
In order for this system to have non-zero solutions for ${\bf f}:=(F_A,F_B,F_C)^T$ 
we require that the matrix ${\bf M}$ is singular (det${\bf M}=0$) which implies 
\begin{align} 
0 = \; & (\gamma+\lambda - W ) \left[ \gamma \lambda \Theta + 2\gamma^2 
+ 2 \lambda^2 - 2\gamma \lambda  - \ (\lambda+\gamma) W    -  W^2 \right] ,  
\lbl{w-cubic} \\ 
W = \; & \omega^2 - 2\lambda - 2\gamma - \Omega^2 , \nn \\ 
\Theta(k,l) = \; & 2 \cos(2k) + 4 \cos k \cos l h = 4 \cos k (\cos k + \cos lh) - 2 . 
\lbl{THdef} \end{align}
The three solutions for $\omega^2$ are a `flat' band, that is, the frequencies 
are independent of the wavenumbers $k,l$, the frequencies being given by 
\beq
\omega_3^2(k,l) =  \Omega^2 + 3\lambda + 3 \gamma , \lbl{w-flat}
\eeq
together with acoustic (lower) and optical (upper) bands given by  
\beq
\omega_{1,2}^2(k,l) =  \Omega^2  + \mfrac{3}{2}\lambda 
+ \mfrac{3}{2} \gamma \pm \half \sqrt{ 9 \lambda^2 + 9 \gamma^2 
- 6\gamma\lambda + 4 \gamma\lambda \Theta(k,l) } . 
\lbl{w12}\eeq
For most values of $(k,l)$, there are three distinct values for $\omega^2$. 
However, for some $(k,l)$, the system is degenerate, that is, two surfaces 
meet.  This could be where the acoustic and optical bands meet and 
form Dirac points, or where the flat band meets one of the acoustic/optical 
branches.  At such points the matrix equation (\ref{mf-equal-zero}) is 
doubly degenerate, and so we need to write ${\bf f}$ in terms of 
two eigenvectors and then higher terms in the  expansion ansatz 
will also depend on two leading-order envelope functions. 
The only conditions where all three surfaces can meet is if $\lambda=\gamma=0$, 
which means there is no nearest-neighbour interactions in the lattice; 
we ignore this possibility. 

\begin{figure}[htb] 
\includegraphics[width=160mm,height=75mm]{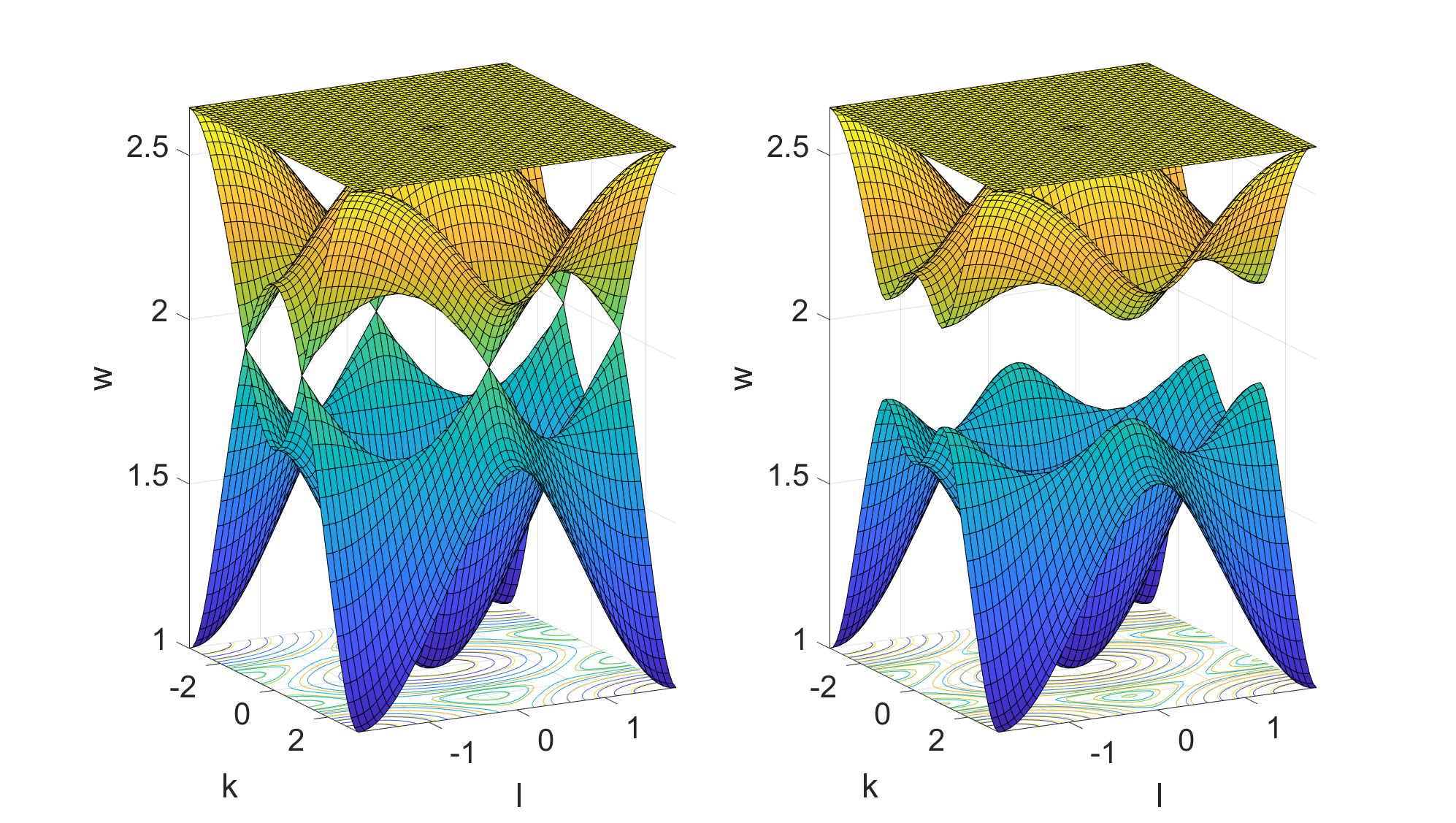}
\caption{Illustration of the dispersion surfaces for the cases 
$\Omega=1=\lambda=\gamma$ (left) and, (right), the case 
$\Omega=1$, $\gamma=1.2$, $\lambda=0.8$. The former cases exhibits 
Dirac points, whereas the latter does not.  Both cases have tangential 
meeting of the flat mode with the optical mode at $k=l=0$.  
(In colour on-line) \lbl{surf-fig} }
\end{figure}

Figure \ref{surf-fig} shows the flat band and the two wave-number-dependent 
surfaces (acoustic and optical).  Note that in this case the flat band lies 
above the other two.  In the case $\gamma=\lambda$ we have Dirac 
points where the acoustic and optical surfaces meet and form cone-shaped 
singularity, whereas in the more general case $\gamma\neq\lambda$, 
we have smooth surfaces and stationary modes at these points 
and a gap between the two surfaces.  
In figure \ref{w2-fig} we present contour plots of the acoustic and optical 
surfaces, which shows the hexagonal symmetry of the system. 

\begin{figure}[htb] 
\includegraphics[scale=0.55]{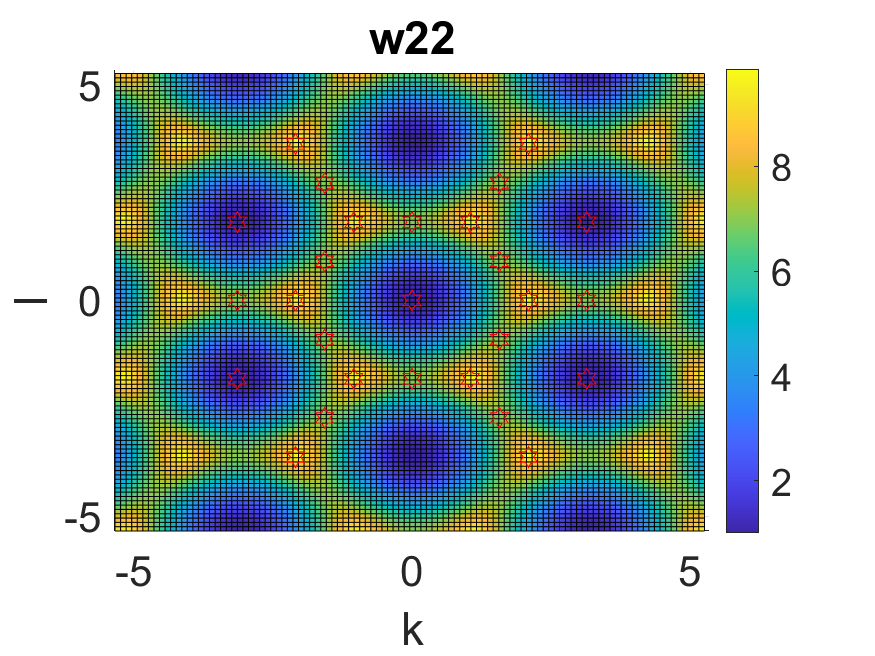}
\includegraphics[scale=0.55]{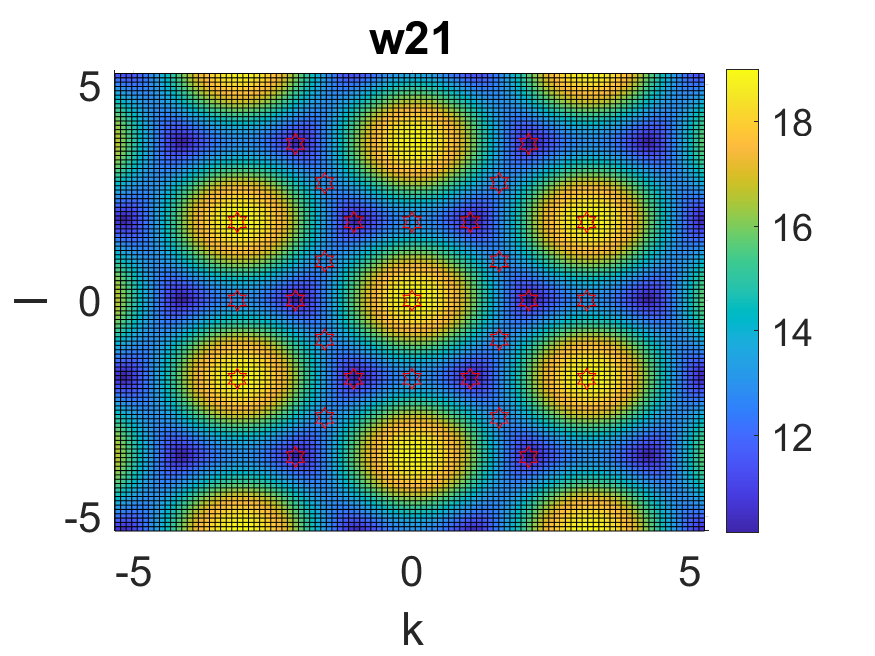}
\caption{Dispersion relations for: left: acoustic mode, $\omega_{ac}^2(k,l)$, 
right, optical $\omega_{opt}^2(k,l)$; for $\gamma=\lambda=3$, $\Omega=1$. 
This is the case where Dirac points occur, these are located at the corners 
of the hexagons (in the case $\lambda \neq \gamma$, these points 
correspond to the acoustic maxima and optical minima).   
The centres of the hexagons correspond to the global maxima 
and global minima points, the saddle points occur in the centres 
of the hexagons' edges.  See Table \ref{crit-tab},  
(in colour, in on-line version). \lbl{w2-fig} }
\end{figure}

We now aim to find stationary points of the dispersion relation $\omega(k,l)$, 
which correspond to stationary points of $\Theta(k,l)$ 
where we expect to find stationary breather-modes, thus we solve 
$\partial\Theta/\partial k=0=\partial \Theta/\partial l$, which gives  
\beq
\sin k \left[ 2 \cos k + \cos lh \right] =0 , \qquad \cos k \sin l h =0 . 
\eeq
The solutions of this pair of coupled equations can be obtained as 
\begin{description}
\item{(i)} $\sin l h = 0$ and $\sin k =0$ so $(k,l) = (m\pi , n\pi/h)$;   
\item{(ii)} $\sin l h =0 $ and $ \cos k = - \half \cos lh$ so $l = n \pi / h$ 
and then either: 
\begin{description}
\item{(ii)(a)} $l = 2 n \pi / h$ and $\cos k =- \half$ so 
$(k,l)=(\pm 2\pi/3 + 2m\pi, 2n\pi/h)$;  or 
\item{(ii)(b)} $l=(2n+1)\pi/h)$ and $\cos k = +\half$ so 
$(k,l) = ( \pm \pi/3 +2m\pi, (2n+1)\pi/h)$; 
\end{description}
\item{(iii)} $\cos k=0$ so that  $k=(2n+1)\pi/2$ and $\cos lh=0$ so 
$(k,l) = ( (2n+1)\pi/2 , (2m+1)\pi/2h)$; 
\item{(iv)} $\cos k =0$ so that $k=(2n+1)\pi/2$ and $\sin k=0$, 
which is self-contradictory. 
\end{description}
These results are summarised in Table \ref{crit-tab}, 
they are the maxima and minima of each surface of the dispersion 
relation. 

\begin{table}[ht] 
\begin{center}
\begin{tabular}{ | c | l | c | c | c | } \hline 
Case & Description & e.g.~$(k,l)$ &  $\Theta(k,l)$ & $\omega^2$ \\ \hline 
1 & Global min (ac) & $(0,0)$  & $6$ & $\Omega^2$  \\
2 & Acoustic saddle & $(0,\pm\pi/h)$  & $-2$ & $\Omega^2 
\!+\! \mfrac{3}{2}\left(\lambda+\gamma+\sqrt{\lambda^2\!+\!\gamma^2
\!-\!\mfrac{14}{9}\lambda\gamma}\,\right) $  \\
3 & Acoustic max $(\gamma\neq\lambda)$ & $(\pm\mfrac{2}{3}\pi,0)$ & $-3$ & $\Omega^2 \!+\! 
\mfrac{3}{2} (\lambda\!+\!\gamma\!-\!|\lambda\!-\!\gamma|) 
= \Omega^2 \!+\! 3\min\{\gamma,\lambda\}$ \\  \hline 
4 & Dirac (if $\lambda\!=\!\gamma$ then & $(\pm\mfrac{2}{3}\pi,0)$ & $-3$ & 
$\Omega^2 + \mfrac{3}{2}(\gamma\!+\!\lambda) 
= \Omega^2\!+\!3\gamma=\Omega^2\!+\!3\lambda$ \\
 & ac-max $\equiv$ opt-min) &   & &  \\ \hline 
5 & Optical min ($\gamma\neq\lambda$) & $(\pm\mfrac{2}{3}\pi,0)$ & $-3$ & $\Omega^2 \!+\! 
 \mfrac{3}{2} (\lambda\!+\!\gamma \!+\! |\lambda\!-\!\gamma|) 
 = \Omega^2 \!+\! 3\max\{\gamma,\lambda\}$ \\ 
6 & Optical saddle &  $(0,\pm\pi/h)$ & $-2$ & $\Omega^2 
+ \mfrac{3}{2}\left(\lambda+\gamma-\sqrt{ \lambda^2+\gamma^2
-\mfrac{14}{9}\lambda\gamma } \, \right) $ \\
7 & Global max (opt) &  $(0,0)$ & $6$ & $\Omega^2 + 3\gamma+3\lambda$ \\ \hline 
8 & Flat band (upper) & any $(k,l)$ & n/a & $\Omega^2+3\gamma+3\lambda$  \\ \hline 
\end{tabular}
\end{center}
\caption{ Summary of critical points, in order of increasing frequency $\omega(k,l)$. 
\lbl{crit-tab} }
\end{table}

\subsection{{\bf Asymptotic approximation of general breather modes}} 

\blue{Whilst some properties of the kagome lattice can be deduced 
from the dispersion relation, we now aim to determine in more 
detail how disturbances propagate through the lattice. 
We aim to find the equations which govern the shape of 
envelope solutions for small amplitude weakly nonlinear 
oscillations, using an asymptotic ansatz.} 
We seek breather modes with leading order envelope given by  
\begin{align}
q_{m,n}^K(t) = & \; \ep \eee{} F_K(x,y,\tau,T)  + \ep^2 ( \eee{} G_K + \eee{2} I_K + J_K) 
\nn \\ & + \ep^3 \eee{} H_K + \ldots + c.c. , \qquad K  = \{ A,B,C \} , 
\lbl{ansatz} \end{align}
where $F_K$ are not constants, rather, they are functions which 
vary slowly over space and time, depending on the slow variables 
\beq
x= \ep m , \quad y = \ep n h , \quad \tau = \ep t , \quad 
T = \ep^2 t , \quad h = \sqrt{3} . \lbl{xytauT}
\eeq
The quantities $G_K,H_K,I_K,J_K$ are also functions of $x,y,\tau,T$ 
and determine higher order correction terms. 
We substitute the ansatz (\ref{ansatz}) into the governing equations 
(\ref{qAddot})--(\ref{qCddot}) and expand, equating terms of equal order 
in $\epsilon$ to generate a hierarchy of solvable equations.   
Since we only consider symmetric potentials, no second harmonics 
are generated, and so we can ignore $I_K,J_K$;  if there were quadratic 
nonlinearities in (\ref{qAddot})--(\ref{qCddot}), then $I_K,J_K\neq0$ 
and we would have to find $I_K,J_K$ by considering the equations 
of motion at $\mathcal{O}(\ep^2)$ and $\mathcal{O}(\ep^2\eee{2})$. 
The presence of $\beta\neq0$ together with $q_{m,n}=\mathcal{O}(\ep)$ 
means that at $\mathcal{O}(\ep^3)$, we obtain correction terms 
due to the weakly nonlinear terms. 

\blue{In the case of $\gamma \neq \lambda$, there are no Dirac 
points in the dispersion relation; instead, there is a gap between 
the maximum of the acoustic band, Case 3, and the minimum 
of the optical band, Case 5.   In this gap, we expect to see 
breathers, and since this case has been considered in detail 
by Hofstrand \cite{Hof}, we focus on other cases below. }

\subsubsection{The leading order solution, $\mathcal{O}(\ep\eee{})$} 

The terms of $\mathcal{O}(\ep)$ in (\ref{qAddot})--(\ref{qCddot}) 
yield a system of linear equations for $F_A,F_B,F_C$ which is 
still given by (\ref{mf-equal-zero}).  
In order for ${\bf Mf=0}$, we still need det(${\bf M}$)=0 
which gives the dispersion relations (\ref{w-flat}) and (\ref{w12}). 
In addition, to solve the equations at higher order, 
we need to know the form of solutions for ${\bf f} = (F_A,F_B,F_C)^T$, 
which are given by $F(x,y,\tau,T){\bf k}$ where ${\bf k}$ 
is the kernel of ${\bf M}$. 

\subsubsection{First correction terms, $\mathcal{O}(\ep^2\eee{})$} 

\blue{To obtain the first correction terms, we expand 
the difference terms using a Taylor series in $x$ and $y$; 
from (\ref{ansatz}) for $q_{m+1,n+1}^K$, writing 
$\theta = k m + hln-\omega t$ we obtain 
\begin{align}
q_{m+1,n+1}^K = \; & 
\ep \ee^{i\theta+ik+ilh} F_K(x\!+\!\ep,y\!+\!\ep h,\tau,T)  
+\ep^2\ee^{i\theta+ik+ilh} G_K(x\!+\!\ep,y\!+\!\ep h,\tau,T) 
+ \ep^2 J_K(x\!+\!\ep,y\!+\!\ep h,\tau,T)) 
\nn\\ & 
+ \ep^2 \ee^{2i\theta+2ik+2ilh} I_K(x\!+\!\ep,y\!+\!\ep h,\tau,T) 
+ \ep^3 \ee^{i\theta+ik+ilh} H_K + \ldots + c.c. , \nn \\ 
= \; & 
\ep \ee^{i\theta+ik+ilh} F_K 
+ \ep^2 \ee^{i\theta+ik+ilh} F_{K,x}
+ \ep^2 h \ee^{i\theta+ik+ilh} F_{K,y}  
+ \ep^2 \ee^{i\theta+ik+ilh} G_K 
+ \ep^2 \ee^{2i\theta+2ik+2ilh} I_K  + \ep^2 J_K  \nn \\ & 
+\half \ep^3 \ee^{i\theta+ik+ilh} F_{K,xx}
+\mfrac{3}{2}\ep^3 \ee^{i\theta+ik+ilh} F_{K,yy}
+\ep^3 h \ee^{i\theta+ik+ilh} F_{K,xy} 
+ \ep^3 J_{K,x} + \ep^3 h J_{K,y}  
+ \ep^3 \ee^{i\theta+ik+ilh} H_K
\nn \\ & 
+ \ep^3 \ee^{i\theta+ik+ilh} G_{K,x} 
+ \ep^3 h \ee^{i\theta+ik+ilh} G_{K,y} 
+ \ep^3 \ee^{2i\theta+2ik+2ilh} I_{K,x}  
+ \ep^3 h \ee^{2i\theta+2ik+2ilh} I_{K,y} + \ldots + c.c.
\lbl{expansion} \end{align}
with similar formulae holding for $q_{m\pm1,n\pm1}$ 
and $q_{m\pm2,n}$. 
Note that there is only one $\mathcal{O}(\ep)$ term, which is of the 
form already included in the leading-order analysis above.   The 
remaining terms are all $\mathcal{O}(\ep^2)$, or $\mathcal{O}(\ep^3)$.  
Of the $\mathcal{O}(\ep^3)$ terms, only those with the wavenumber
and frequency $\eee{}$ are relevant, and these are considered below, 
in Section \ref{ep3e1sec}.  }

 Hence, at second order, we find 
\begin{align}
& (\omega^2-\Omega^2-2\gamma-2\lambda) G_A 
+ (\gamma +\lambda \ee^{ik+ilh} ) G_B 
+ ( \gamma +\lambda \ee^{-ik+ilh}) G_C 
\nn \\ & \qquad 
=  -2 i \omega F_{A,\tau} - \lambda\ee^{ik+ilh} F_{B,x} 
- \lambda\ee^{ik+ilh} h F_{B,y} + \lambda \ee^{-ik+ilh} F_{C,x} 
- \lambda \ee^{-ik+ilh} h F_{C,y} , \lbl{genGAeq}
\\[2ex] 
& (\gamma +\lambda \ee^{-ik-ilh} ) G_A 
+ (\omega^2-\Omega^2-2\gamma-2\lambda) G_B 
+ (\gamma +\lambda \ee^{-2ik} ) G_C 
\nn \\ & \qquad 
=  -2 i \omega F_{B,\tau} + \lambda \ee^{-ik-ilh} F_{A,x} 
+ \lambda \ee^{-ik-ilh} h F_{A,y}  + 2 \lambda \ee^{-2ik} F_{C,x} , 
\\[2ex] 
& (\gamma + \lambda \ee^{ik-ilh} ) G_A 
+ (\gamma+\lambda\ee^{2ik} ) G_B 
+ (\omega^2-\Omega^2-2\gamma-2\lambda )  G_C 
\nn \\ & \qquad 
=  -2 i \omega F_{C,\tau} - \lambda \ee^{ik-ilh} F_{A,x} 
+ \lambda h \ee^{ik-ilh} F_{A,y} - 2\lambda \ee^{2ik} F_{B,x} .  
\lbl{genGCeq} \end{align}
This has the form ${\bf M}{\bf g} = {\bf b}$, where ${\bf g}
= (G_A,G_B,G_C)^T$ and ${\bf b}$ is given by the right-hand 
sides of (\ref{genGAeq})--(\ref{genGCeq}).  
Since ${\bf M}$ is singular, we have a consistency condition 
on ${\bf b}$ in order for a solution to exist, 
\blue{which is the Fredholm alternative. 
In this case, since ${\bf M}$ is Hermitian, its eigenvalues 
are real, and its eigenvectors are orthogonal with respect 
to the inner product $\langle {\bf e}_1 , {\bf e}_2 \rangle 
= {\bf e}_1^* \cdot {\bf e}_2$.  Since ${\bf M}{\bf f}={\bf 0}$, 
in order to solve ${\bf Mg} = {\bf b}$ we require 
$\langle {\bf f} , {\bf b} \rangle={\bf f}^* \cdot {\bf b}=0$ 
in order for there to be solutions for ${\bf g}$}. 
Once this condition has been met, the solution for ${\bf g}$ has 
to be found, and this can include an arbitrary component of ${\bf f}$, 
which is the kernel of ${\bf M}$; we set this component to zero 
without losing any generality, since that mode is accounted 
for in the leading order solution. 

\subsubsection{Derivation of NLS from $\mathcal{O}(\ep^3\eee{})$ \lbl{ep3e1sec}} 

\blue{Expanding (\ref{qAddot})--(\ref{qCddot}) after 
substituting in the ansatz (\ref{ansatz}) using (\ref{expansion}), 
and retaining terms at $\mathcal{O}(\ep^3)$ with 
frequency $\eee{}$, we find that}
the second order corrections, $H_K$, must satisfy 
\begin{align}
& (\omega^2-\Omega^2-2\gamma-2\lambda ) H_A 
+ (\gamma + \lambda \ee^{ik+ilh} ) H_B 
+(\gamma + \lambda \ee^{-ik+ilh} ) H_C 
\nn \\ & \qquad 
= - 2 i \omega G_{A,\tau} + F_{A,\tau\tau} - 2 i \omega F_{A,T} 
+ 3 \beta |F_A|^2F_A   
+\lambda \ee^{-ik+ilh} G_{C,x}
- \lambda h\ee^{-ik+ilh} G_{C,y} \nn \\ & \qquad\qquad 
-\half \lambda \ee^{-ik+ilh} F_{C,xx}
-\mfrac{3}{2}\lambda \ee^{-ik+ilh} F_{C,yy}
+\lambda h\ee^{-ik+ilh} F_{C,xy}
-\lambda \ee^{ik+ilh} G_{B,x}
-\lambda h\ee^{ik+ilh} G_{B,y} \nn \\ & \qquad\qquad 
-\half \lambda \ee^{ik+ilh} F_{B,xx}
-\mfrac{3}{2} \lambda \ee^{ik+ilh} F_{B,yy}
-\lambda h\ee^{ik+ilh} F_{B,xy} , \lbl{genHAeq}
\\[2ex] 
& (\gamma + \lambda \ee^{-ik-ilh} ) H_A 
+ (\omega^2-\Omega^2-2\gamma-2\lambda) H_B 
+ (\gamma + \lambda \ee^{-2ik} ) H_C 
\nn \\ & \qquad 
= 2 i \omega G_{B,\tau} + F_{B,\tau\tau} - 2 i \omega F_{B,T} 
+ 3 \beta |F_B|^2 F_B +\lambda \ee^{-ik-ilh} G_{A,x}
+\lambda h\ee^{-ik-ilh} G_{A,y} \nn \\ & \qquad\qquad 
-\half \lambda \ee^{-ik-ilh} F_{A,xx}
-\mfrac{3}{2} \lambda \ee^{-ik-ilh} F_{A,yy}
-\lambda h \ee^{-ik-ilh} F_{A,xy}
+2 \lambda \ee^{-2ik} G_{C,x}
- 2 \lambda \ee^{-2ik} F_{C,xx} , 
\\[2ex]
& (\gamma + \lambda \ee^{ik-ilh}) H_A 
+ ( \gamma +\lambda \ee^{2ik} ) H_B 
+ (\omega^2-\Omega^2-2\gamma-2\lambda )  H_C 
\nn \\ & \qquad 
=  - 2 i \omega G_{C,\tau} + F_{C,\tau\tau} - 2 i \omega F_{C,T} 
+ 3 \beta |F_C|^2 F_C
- 2 \lambda \ee^{2ik} G_{B,x}
- 2 \lambda \ee^{2ik} F_{B,xx} \nn \\ & \qquad \qquad 
-  \lambda \ee^{ik-ilh} G_{A,x}
+ \lambda h \ee^{ik-ilh} G_{A,y}
- \half \lambda \ee^{ik-ilh} F_{A,xx}
- \mfrac{3}{2}\lambda \ee^{ik-ilh} F_{A,yy}
+ h \lambda \ee^{ik-ilh} F_{A,xy} . \lbl{genHCeq}
\end{align}
\blue{The process of obtaining systems of equations such as this, 
is also known as complexification averaging. } 
At this stage, we do not need to find $H_A,H_B,H_C$, 
we just apply the solvability condition on the right-hand 
sides in order to determine the equation(s) for the leading 
order terms $F_A,F_B,F_C$. 
Typically, this gives an NLS equation, since we have second 
spatial derivatives, first derivatives in time with a pure imaginary 
prefactor and cubic nonlinear terms. 
Below, we consider various special cases. 

\subsection{{\bf Case 1: the global minimum}} \lbl{min-sec}

The global minimum of the frequencies $\omega$ occurs on the 
acoustic branch, at $(k,l)=(0,0)$ where $\Theta=6$ and $\omega=\Omega$, 
from (\ref{mf-equal-zero}) we have 
\begin{align} 
{\bf M f} =  (\gamma+\lambda) \mat{-2}{1}{1 }{ 1}{-2}{1 }{ 1}{1}{-2} 
\vect{F_A}{F_B}{F_C} = \vect{0}{0}{0}\!. 
\lbl{c1-Mf}
\end{align}
Hence we write the leading order solution as ${\bf f} = 
(F_A,F_B,F_C)^T=F (1,1,1)^T$ since the kernel of ${\bf M}$ 
\blue{given by the span of} $(1,1,1)^T$; 
here $F=F(x,y,\tau,T)$.
To solve the general singular system ${\bf M v}={\bf b}$, 
since the range of ${\bf M}$ is $\kappa_1 (-2,1,1)^T + \kappa_2 (1,-2,1)^T$ 
for some $\kappa_1,\kappa_2$, which has normal ${\bf n} = (1,1,1)$, the 
consistency condition for the existence of a solution is ${\bf b} \cdot (1,1,1)^T=0$. 

\subsubsection{Case 1 correction terms, from $\mathcal{O}(\ep^2\eee{})$} 

Considering the largest of the correction terms, 
\blue{putting $k=0=l$ and $\omega^2=\Omega^2$ in } 
(\ref{genGAeq})--(\ref{genGCeq}) we obtain the equations 
\begin{align}
(\gamma +\lambda  ) (-2G_A +G_B +G_C) = &\; 
-2 i \omega F_{A,\tau} - \lambda F_{B,x} - \lambda  h F_{B,y} 
+ \lambda F_{C,x} - \lambda  h F_{C,y} , \lbl{c1-GA-eq} \\[1ex] 
(\gamma +\lambda  ) (G_A - 2G_B+G_C) = &\; 
-2 i \omega F_{B,\tau} + \lambda F_{A,x} 
+ \lambda  h F_{A,y} + 2 \lambda  F_{C,x} ,  \\[1ex] 
 (\gamma + \lambda  ) (G_A+G_B-2 G_C) = &\; 
 -2 i \omega F_{C,\tau} - \lambda F_{A,x} + \lambda h  F_{A,y} 
 - 2\lambda F_{B,x} . \lbl{c1-GC-eq}
\end{align}
This has the form ${\bf M}{\bf g} = {\bf b}$ with ${\bf g}=(G_A,G_B,G_C)$ 
and since ${\bf M}$ is singular, we require a consistency condition 
to be satisfied for solutions to exist, which implies that $F_{\tau}=0$, \blue{that is, we rewrite $F(x,y,\tau,T) = F(x,y,T)$}
so the mode is stationary on the intermediary timescale $\tau$, \blue{that is, $F$ is independent of $\tau$ , 
although it may still evolve on the very long timescale, $T$}. 
Noting that $F_A=F_B=F_C=F$, the right-hand sides of 
(\ref{c1-GA-eq})--(\ref{c1-GC-eq}) simplify; we have ${\bf Mg}={\bf b}$ 
with ${\bf b} = \lambda ( -2 h F_y , 3 F_x + h F_y, -3 F_x + h F_y)^T$, 
thus one of the many available solutions for ${\bf g}$ is 
\beq
{\bf g }  = \vect{G_A}{G_B}{G_C}  =  \frac{\lambda}{3(\lambda+\gamma)}  
\vect{ 2 h F_y }{ -h F_y - 3 F_x }{ - h F_y + 3 F_x } .  \lbl{c1-Gsol}
\eeq
Any vector of the form $G_0(x,y,T)(1,1,1)^T$ could be added to this, 
but since a vector of this form is used as the leading order solution 
we can neglect $G_0$ at this order. 

\subsubsection{Case 1, second-order correction terms, from $\mathcal{O}(\ep^3\eee{})$} 

At third order, from (\ref{genHAeq})--(\ref{genHCeq}), we obtain 
a system of equations for the terms $(H_A,H_B,H_C)^T$ 
\begin{align}
(\gamma \!+\! \lambda ) (-2H_A\!+\!H_B\!+\!H_C)  = & \; 
- 2 i \omega G_{A,\tau} + F_{A,\tau\tau} - 2 i \omega F_{A,T} 
+ 3 \beta |F_A|^2F_A  +\lambda  G_{C,x} - \lambda h G_{C,y} 
-\lambda G_{B,x} \nn  \\ &   -\lambda h  G_{B,y} 
-\half \lambda F_{C,xx} -\mfrac{3}{2}\lambda  F_{C,yy}
+\lambda h  F_{C,xy} -\half \lambda  F_{B,xx}
-\mfrac{3}{2} \lambda F_{B,yy} -\lambda h  F_{B,xy} , 
\nn \\ & \lbl{c1-HA} \\[1ex] 
 (\gamma \!+\! \lambda ) ( H_A \!-\! 2H_B \!+\! H_C) = & \; 
- 2 i \omega G_{B,\tau} + F_{B,\tau\tau} - 2 i \omega F_{B,T} 
+ 3 \beta |F_B|^2 F_B +\lambda  G_{A,x}
+\lambda h  G_{A,y} \nn \\ & 
-\half \lambda  F_{A,xx} -\mfrac{3}{2} \lambda  F_{A,yy}
-\lambda h  F_{A,xy} +2 \lambda G_{C,x} - 2 \lambda F_{C,xx} , 
\\[1ex]
(\gamma \!+\! \lambda ) ( H_A \!+\! H_B \!-\! 2 H_C) = & \; 
 - 2 i \omega G_{C,\tau} + F_{C,\tau\tau} - 2 i \omega F_{C,T} 
+ 3 \beta |F_C|^2 F_C - 2 \lambda G_{B,x} - 2 \lambda F_{B,xx} 
\nn \\ & 
-  \lambda G_{A,x} + \lambda h  G_{A,y} - \half \lambda F_{A,xx}
- \mfrac{3}{2}\lambda F_{A,yy} + h \lambda F_{A,xy} . 
\lbl{c1-HC}  \end{align}
At this stage, we do not need to find ${\bf h}=(H_A,H_B,H_C)^T$,  
we only need the consistency condition to be satisfied, that is, if 
(\ref{c1-HA})--(\ref{c1-HC}) is written as ${\bf Mh}= {\bf b}$, 
with ${\bf M}$ given by (\ref{c1-Mf}) then ${\bf b} \cdot (1,1,1)^T=0$. 
Using the solutions (\ref{c1-Gsol}) implies 
\begin{align}
i \omega F_T = & \; \mfrac{3}{2} \beta |F|^2 F - 
\frac{\lambda\gamma}{\lambda+\gamma} ( F_{xx} + F_{yy} ) .  
\lbl{NLSFmin} \end{align}
This is a standard NLS equation in 2+1 dimensions, and has 
solitary wave solution given by the Townes soliton \cite{Townes},  
in the case $\beta\gamma\lambda(\lambda+\gamma)<0$, 
\blue{as illustrated in Figure \ref{Townes-fig}. }  
\blue{Since we expect $\gamma,\lambda>0$ in (\ref{ham}), 
this corresponds to $\beta<0$, 
which means that the nonlinearity is softening, 
that is, a doubling of the displacement $q$ 
results in a restoring force which is less than doubled. } 

\begin{figure}[hbtb] 
\includegraphics[scale=0.4]{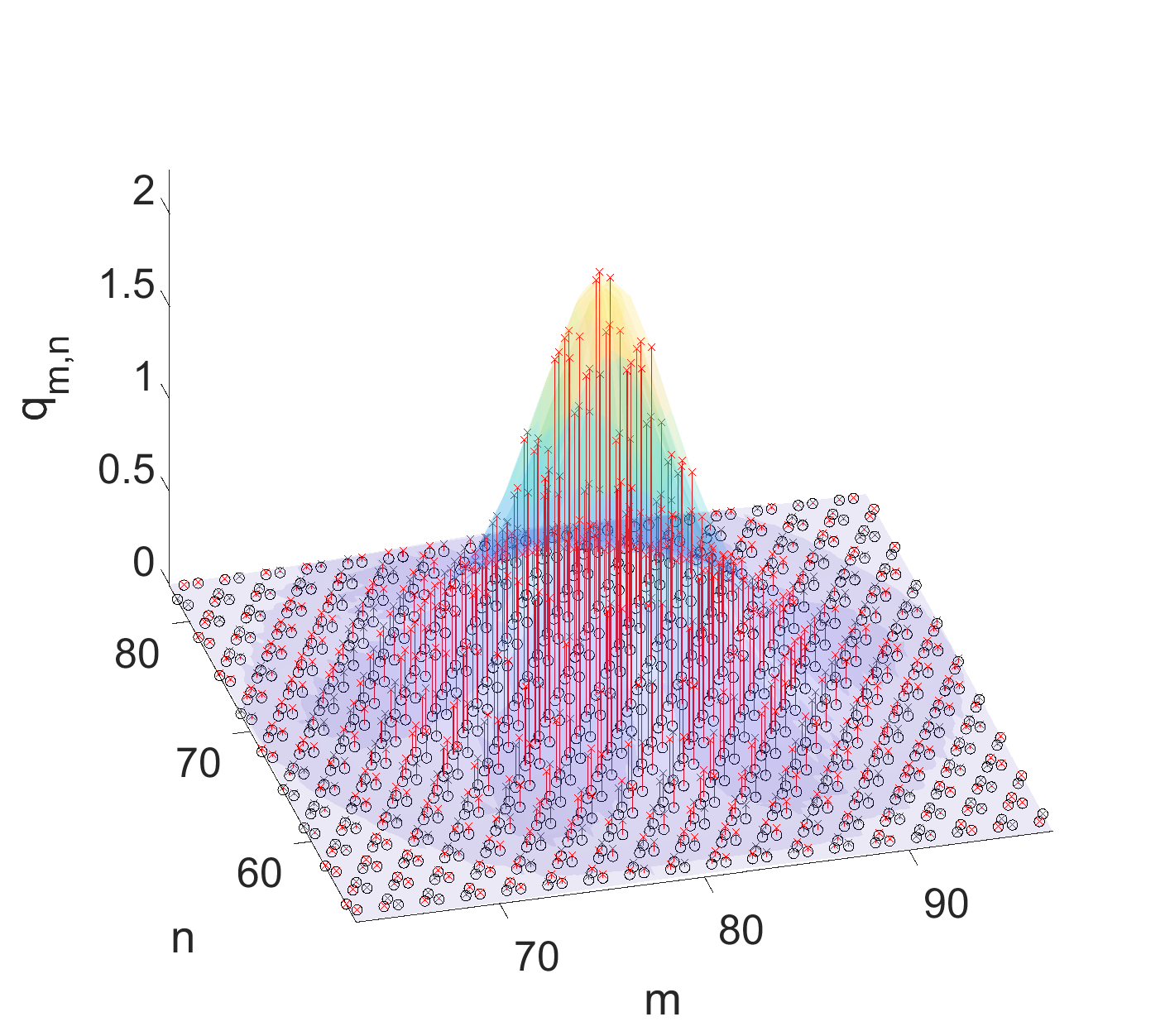} 
\caption{ Illustration of the Townes soliton solution on 
the kagome lattice, red crosses indicate the solution 
$q_{m,n}=2\ep \cos(t(\Omega-\ep^2\tilde\Omega) \Psi(r)$, 
which is plotted against lattice nodes $(m,n)$ (where $m+n$ even), 
with $r^2 = \ep^2((m-M/2)^2 + 3(n-N/2)^2$ evaluated at $t=0$,   
for the parameter values $\omega=\Omega=2$, 
$\gamma=\lambda=1/2$,  $\widetilde\Omega=-1/2$, 
$\beta=-0.1$,  Black circles indicate the 
equilibrium configuration $q_{m,n}=0$, 
red lines join the zero state to the displaced red crosses 
(in colour in on-line version).   \lbl{Townes-fig} }
\end{figure}

If we transform to a moving coordinate frame {\em via} 
$z=x-UT$ and $F(x,y,T)=\widetilde{F}(z,y,T)\ee^{iJz-iS T}$, 
then we find $\widetilde F, J, L$ are given by 
\beq 
i \omega \widetilde F_T = \frac{3}{2} \beta | \widetilde F |^2 \widetilde F 
- \frac{\lambda\gamma}{\lambda+\gamma} ( \widetilde F_{zz} + \widetilde F_{yy} ) , \qquad 
J = \frac{\omega U(\lambda+\gamma)}{2\lambda\gamma} , \qquad 
S = - \frac{ \omega U^2 (\lambda+\gamma)}{4 \lambda\gamma} .  \lbl{C1nls}
\eeq 
Since the PDE for $\widetilde F$ is the same as (\ref{NLSFmin}), 
we have a stationary Townes solitons for $\widetilde{F}$, which 
corresponds to a moving breather for $F(x,y,T)$.
Thus we can obtain slowly moving breathers near the minimum 
of the dispersion relation. 
\blue{However, in the classic NLS equation, this soliton is unstable 
due to the Vakhitov-Kolokolov instability \cite{kolo,vk}. 
The wave can either spread out indefinitely, or collapse to a point, 
as all energy focuses on a single point.   In a lattice 
this latter mechanism cannot occur, and higher order 
derivatives in the Taylor expansions, and higher-order nonlinearities 
may stabilise the mode as suggested by the work of Karpman 
\cite{karpman}, and Davydova \etal\ \cite{davydova} 
on generalisations of the NLS equation.} 

Cases 2,3,5,6 follow similar reduction to the above form, namely 
the matrix ${\bf M}$ has a one-dimensional kernel, 
so  $F_A,F_B,F_C$ can all be written in terms of a single quantity, $F$; 
the range of ${\bf M}$ is two-dimensional, so that 
at second order, there is a single condition, 
and $G_A,G_B,G_C$ can be found in terms of the first derivatives of $F$; 
and the single condition applied again at third-order gives a single NLS 
equation in 2+1 dimensions. 
Cases 4 and 7-8 are, however, different, and so we analyse these 
in the following subsections. 

\subsection{{\bf Case 4 Dirac point, $(k,l)=(\mfrac{2}{3}\pi,0)$ with $\lambda=\gamma$}} \lbl{Dirac-sec}

Dirac points only occur if $\gamma=\lambda$, hence this 
will be assumed for the remainder of this section.  
If this condition fails, then there is a gap between the acoustic 
and optical modes, and we have separate modes on each 
surface, each satisfying a single 2D NLS equation 
similar to (\ref{C1nls}). 
At a Dirac point, the two surfaces meet at a single point, 
and the matrix in the leading order problem becomes 
doubly degenerate, having a two-dimensional kernel. 
To illustrate this case, we consider the 
example $(k,l)=(\pm\mfrac{2}{3}\pi,0)$, giving $\Theta=-3$ 
$\omega^2 = \Omega^2 + 3\lambda$, and 
\begin{align}
{\bf M v} := \half \gamma \mat{-2} {1+i\sqrt{3} }{1-i\sqrt{3} }
{1- i\sqrt{3}}{ -2 }{ 1+i \sqrt{3} }{ 1+i\sqrt{3} }{ 1-i\sqrt{3} }{-2} 
\!\vect{F_A}{F_B}{F_C} \!=\! \vect{0}{0}{0} . 
\end{align}
We write the leading order general solution as  
\beq 
\vect{F_A}{F_B}{F_C} = F(x,y,\tau,T) \vect{2}{2}{2} 
+ P(x,y,\tau,T) \vect{ -2}{1+ih }{1-ih} ,   \lbl{Dirac-F}
\eeq
the vectors being chosen to have equal magnitudes.

\subsubsection{Case 4: first correction terms--from $\mathcal{O}(\ep^2\eee{})$} 

From (\ref{genGAeq})--(\ref{genGCeq}) we have 
\begin{align} 
{\bf M} \vect{G_A}{G_B}{G_C} = \vect{
- 2 i \omega F_{A,\tau} - \lambda\ee^{ik+ilh} F_{B,x} 
- \lambda\ee^{ik+ilh} h F_{B,y} + \lambda \ee^{-ik+ilh} F_{C,x} 
- \lambda \ee^{-ik+ilh} h F_{C,y}
}{ 
 - 2 i \omega F_{B,\tau} + \lambda \ee^{-ik-ilh} F_{A,x} 
+ \lambda \ee^{-ik-ilh} h F_{A,y}  + 2 \lambda \ee^{-2ik} F_{C,x} 
}{ 
 - 2 i \omega F_{C,\tau} - \lambda \ee^{ik-ilh} F_{A,x} 
+ \lambda h \ee^{ik-ilh} F_{A,y} - 2\lambda \ee^{2ik} F_{B,x}  
} =: {\bf b} .   \lbl{Dirac2b}
\end{align}
Since the general system ${\bf M v}={\bf b}\neq{\bf 0}$ is 
singular, with the range of ${\bf M}$ being one-dimensional, 
we require ${\bf b} = \kappa_0 (-2, 1-i\sqrt{3}, 1+i\sqrt{3})^T$, 
for some $\kappa_0$. This condition is equivalent to the 
{\em two} orthogonality conditions 
\beq
{\bf b} \cdot \vect{1}{1}{1} = 0 = {\bf b} \cdot \vect{-2}{1-ih}{1+ih} . 
\lbl{Dirac-cond} \eeq
Combining (\ref{Dirac-F}) and (\ref{Dirac-cond}) we obtain 
a system of two coupled first-order linear PDEs 
\begin{align}
0 = \; & -2 i \omega F_\tau + \mfrac{2}{3}  i \lambda h F_x + 
i\lambda h P_x +  \lambda h P_y  , \nn \\ 
0 = \; & -2 i \omega P_\tau +  i \lambda h F_x 
	+ \mfrac{2}{3} i \lambda h P_x - \lambda h F_y , 
\end{align} 
which can be simplified by transforming to a travelling wave 
coordinate $z=x-U\tau$, leading to 
\beq
i P_y = P_z + B F_z , \qquad 
- i F_y = F_z + B P_z , \qquad 
\qomit{ \begin{pmatrix} 1 & B \\ B & 1 \end{pmatrix} 
\begin{pmatrix} F_z \\ P_z \end{pmatrix} = 
i \begin{pmatrix} - F_y \\ P_y \end{pmatrix} , 
\qquad }
B = \frac{2}{3} + \frac{2 \omega U}{\lambda h} . 
\eeq
which reduces to $F_{yy} = (B^2-1)F_{zz}$ which has 
plane wave solutions  of the form $F=F(q)$, 
$q = z \pm y\sqrt{B^2-1}$.  Since these waves are not 
localised in both spatial dimensions ($x,y$) we do not 
pursue this case any further.  
Bahat-Treidel \etal\ \cite{peleg} have considered the 
effects of nonlinearity on Dirac dynamics in a modified 
Schr\"{o}dinger model of a photonic honeycomb lattice. 

\subsection{Case 8: breathers near a generic point on the flat-band \lbl{c8-sec}} 

The analysis of compactly supported solitary waves on lattices 
with flat-bands in their dispersion relation, and the stability of 
waves associated with the flat-band has recently been studied 
by Shi \etal\ \cite{shi}.  They consider NLS equations on 
various lattices (diamond, kagome, checkerboard), 
and establish criteria for the stability of waves 
which are then illustrated via numerical computations 
of the bifurcation diagram.

As noted in Table \ref{crit-tab}, the system (\ref{qAddot})--(\ref{qCddot}) 
has a flat band of linear waves 
whose frequency is $\omega=\pm\sqrt{\Omega^2+3\gamma+3\lambda}$, 
for all wavenumbers, $k,l$.  
The flat band mode has frequency given by (\ref{w-flat}), and the 
amplitudes $F_A,F_B,F_C$ are given by 
\beq
{\bf Mf} = \mat{\lambda\!+\!\gamma }  
{\gamma + \lambda \ee^{ik+ilh} }{\gamma + \lambda \ee^{-ik+ilh} }
{\gamma + \lambda \ee^{-ik-ilh}}{\lambda\!+\!\gamma }
{\gamma + \lambda \ee^{-2ik}}{\gamma + \lambda \ee^{ik-ilh}}
{\gamma + \lambda \ee^{2ik}}{\lambda\!+\!\gamma } 
\!\vect{F_A}{F_B}{F_C} \!=\! \vect{0}{0}{0}\! , 
\eeq
which, in general, implies that ${\bf f}$ is a multiple of the zero eigenvector 
\beq
{\bf f} = F(x,y,\tau,T) \bfk , \quad \bfk = ( 1- \ee^{-2ik} - 2 i \ee^{ilh} \sin(k) \ , \  1 + \ee^{-2ik} 
- 2\ee^{-ik}\cos lh  \ , \ -4 \sin^2\half(k+lh) \ )^T , 
\lbl{bfk-def} \eeq
with $F \neq0$, and more specifically, $F=F(x,y,\tau,T)$, 
and $\bfk$ is the kernel of ${\bf M}$.   
Note that in the limit $(k,l) \rightarrow (0,0)$ this eigenvector 
tends to the zero vector, which highlights the requirement of 
a more general methodology to deal with the special case  
where the flat band meets the optical dispersion surface here; 
see Section \ref{c78-sec} for more details. 

\subsubsection{First-order correction terms, $\mathcal{O}(\ep^2\eee{})$} 

\blue{Applying the consistency condition ${\bf f}^* \cdot {\bf b}=0$ using 
(\ref{Dirac2b}), we obtain the equation $F_\tau=0$, 
and so we write $F(x,y,\tau,T)=F(x,y,T)$. 
Equivalently, we could follow the approach of (\ref{Dirac-sec}) 
and introduce travelling wave coordinates $Z=x-U\tau$, $W=y-V\tau$, 
and seek solutions of the form $F(x,y,\tau,T)=F(Z,W,T)$. 
The consistency condition ${\bf f}^*\cdot {\bf b}$ would then yield 
equations for $U,V$ from the coefficients of $F_Z,F_W$, 
whose solution is $U=0=V$.  
Either approach gives the expected result of waves that move 
with zero speed, which is consistent with the waves speeds $U,V$ 
being proportional to $\partial \omega/\partial k$, 
$\partial \omega /\partial l$, which are both zero, since the band is flat. } 
The equation for ${\bf g}=(G_A,G_B,G_C)^T$ reduces to 
${\bf M}{\bf g}={\bf 0}$ which has the general solution 
${\bf g}=G\bfk$, with $\bfk$ as defined in (\ref{bfk-def}),  but  
we take the solution as ${\bf g}={\bf 0}$, since the leading 
order solution for ${\bf f}=(F_A,F_B,F_C)$ already has a component 
of the form $F\bfk$. Hence we take $G_A=G_B=G_C=0$ in the following. 

\subsubsection{Second-order correction terms, $\mathcal{O}(\ep^3\eee{})$} 

\blue{We again apply the consistency condition ${\bf f}^* \cdot {\bf b} =0$, 
this time to the $\mathcal{O}(\ep^3\eee{})$ equations, that is, 
the right-hand sides of  (\ref{genHAeq})--(\ref{genHCeq}), which yields 
\begin{align}
0 = \; & -8 i w D_T F_T + 6 \beta D_N |F|^2 F + 2 \lambda D_x F_{xx} 
+ 6 \lambda D_y F_{yy} - 4 h\lambda D_m F_{xy} , \lbl{nls8} \\ 
D_T =\; &  1 + \cos^2 lh + \sin^2 k + \sin k \sin lh + 4 \sin^4(\half(k+lh)) - \sin k \sin(2k+lh) - 2 \cos k \cos lh , \nn \\ 
D_x = \; & 2 + 8\cos^2(k) - \cos(hl+3k) + 4\cos^2(lh) - 15\cos(k)\cos(lh) 
+ 5\sin(lh)\sin(k) + 2\cos(2hl+2k) ,  \nn \\ 
D_y = \; & 4\sin^2(k) + \cos(hl+3k) - \cos(lh)\cos(k) + 3\sin(lh)\sin(k) , \nn \\ 
D_m = \; & 1 + 4\sin(lh)\sin(k) + 2\sin^2(lh) - 2\cos^2(k) + \cos(2hl+2k) , \nn \\ 
D_N = \; &  71 + 7\cos(2hl-2k) + 40\cos(2hl+2k)+\cos(4hl+4k) 
- 8\cos(3hl+3k) - 104\cos(hl+k) \nn \\ & 
- 8\cos(hl-k) + 12\cos(hl+3k)-4\cos(3hl-k) 
- 4\cos(3hl+k) + \cos(4lh) - 4\cos(2hl+4k) \nn \\ & 
+ 12\cos(2lh) - 8\cos(2k) 
+ \cos(2hl+6k) + 7\cos(4k) - 4\cos(hl+5k) - 8\cos(hl-3k) . 
\nn \end{align}
The discriminant of this system, $D_i= 48 (D_x D_y - D_m^2)$ 
satisfies $D_i \geq0$ for all $k,l$, so the system is never hyperbolic, 
and is elliptic for almost all $(k,l)$. 
Furthermore, $D_T \leq 0$ and $D_N\geq 0$ for all $(k,l)$, 
so the NLS equation is also always of the focusing form. 
The three quantities $D_T,D_N,D_i$ are all strictly positive 
except along a single curve  in $(k,l)$-space, where all are zero; 
this curve includes the special case $(k,l)=(0,0)$, 
where the upper (optical) surface of the dispersion relation 
is tangential to the flat band, and this special case is analysed 
separately in the next subsection. } 

\subsection{Case 7-8 : Global max/flat-band intersection} \lbl{c78-sec} 

At the wavevector $(k,l)=(0,0)$ the optical branch meets upper 
flat band, and we have $\omega^2 = \Omega^2 + 3\gamma + 
3\lambda$ (since $\Theta=6$).   
Rather than a simple NLS equation as obtained in (\ref{min-sec}), 
we need to consider a more general form of solution. 
At leading order in $\epsilon$, the system can be written as 
\begin{align}
{\bf Mf } = (\lambda+\gamma) \mat{1}{1}{1}{1}{1}{1}{1}{1}{1} 
\vect{F_A}{F_B}{F_C} = \vect{0}{0}{0} . 
\end{align}
Whilst it may be natural to write ${\bf f} = (F_A,F_B,F_C) = F (1,0,-1)
 + P(0,1,-1)$,  we note that due to the symmetry in the problem, 
later calculations are simplified if we instead write 
\beq
{\bf f} = \vect{F_A}{F_B}{F_C} = \vect{2F}{-F-hP}{-F+hP} 
= F \vect{2}{-1}{-1} + h P \vect{0}{-1}{1}.  \lbl{lot78}
\eeq

\subsubsection{Case 7-8 first correction terms $\mathcal{O}(\ep^2\eee{})$} 

We return to the generic case where $\gamma$ is not necessarily equal 
to $\lambda$.  
At the next order, we obtain another highly singular system 
\begin{align}
(\gamma+\lambda) (G_A +  G_B + G_C ) = \; & 
 -2 i \omega F_{A,\tau} - \lambda F_{B,x} 
- \lambda h F_{B,y} + \lambda  F_{C,x} - \lambda  h F_{C,y} , 
\\[1ex] 
(\gamma+\lambda) (G_A +  G_B + G_C ) = \; & 
 -2 i \omega F_{B,\tau} + \lambda  F_{A,x} 
+ \lambda  h F_{A,y}  + 2 \lambda F_{C,x} , 
\\[1ex] 
(\gamma+\lambda) (G_A +  G_B + G_C ) = \; & 
 -2 i \omega F_{C,\tau} - \lambda  F_{A,x} 
+ \lambda h F_{A,y} - 2\lambda F_{B,x}  . 
\end{align}
The consistency condition for the system to have a solution 
is that the three components on the right-hand side are the same, 
that is, ${\bf M}{\bf v} = {\bf q}$ requires ${\bf q} = \alpha (1,1,1)^T$, 
or ${\bf q} \cdot (1,-1,0)^T=0={\bf q} \cdot (1,0,-1)^T$.  
The system is not simplified by transforming to a travelling wave:  
if we were to transform to a moving coordinate frame 
by $z=x-U\tau$, $w=y-V\tau$, then we have $U=V=0$, 
and no $\tau-$dependence in the problem, 
hence we retain $x,y$ as the independent variables. 

The system thus reduces to three copies of the same equation
\beq
(\gamma+\lambda)(G_A+G_B+G_C) = 2 \lambda h (P_x + F_y ) , 
\eeq
which has the solution 
\beq
G_A = G_B = G_C = G = \frac{ 2 h \lambda ( P_x + F_y ) 
}{3(\gamma+\lambda) } . \lbl{78G} 
\eeq
We do not need to include components in the directions 
in (\ref{lot78}) as they can be included in $P,F$. 

\subsubsection{Case 7-8, second order correction terms, $\mathcal{O}(\ep^3\eee{})$} 

At third order in $\epsilon$, we find the equations 
\begin{align}
(\gamma \!+\! \lambda ) (H_A\!+\! H_B \!+\! H_C ) = &  \nn
- \!2 i \omega G_{A,\tau} \!+\! F_{A,\tau\tau} \!-\! 2 i \omega F_{A,T} 
\!+ 3 \beta |F_A|^2F_A  \!+\lambda  G_{C,x}\! - \lambda h  G_{C,y} 
\! -\lambda  G_{B,x} \nn \\ & \! -\!\lambda h G_{B,y}
\!-\!\half \lambda  F_{C,xx} \!-\!\mfrac{3}{2}\lambda F_{C,yy}\!+\!\lambda h  F_{C,xy}
\!-\!\half \lambda F_{B,xx} \!-\!\mfrac{3}{2} \lambda  F_{B,yy} \!-\!\lambda h F_{B,xy} , 
\nn \\  
(\gamma \!+\! \lambda )(H_A\!+\!H_B\!+\! H_C)  = & \; \nn
- 2 i \omega G_{B,\tau} + F_{B,\tau\tau} + 2 i \omega F_{B,T} 
+ 3 \beta |F_B|^2 F_B +\lambda G_{A,x} +\lambda h G_{A,y} \nn \\ &  
-\half \lambda F_{A,xx} -\mfrac{3}{2} \lambda F_{A,yy}
-\lambda h  F_{A,xy} +2 \lambda G_{C,x} - 2 \lambda  F_{C,xx} , 
\nn \\ 
(\gamma \!+\! \lambda)( H_A \!+\!  H_B \!+\! H_C ) = & \; \nn 
- 2 i \omega G_{C,\tau} + F_{C,\tau\tau} - 2 i \omega F_{C,T} 
+ 3 \beta |F_C|^2 F_C- 2 \lambda  G_{B,x} - 2 \lambda  F_{B,xx} \nn \\ &  
-  \lambda  G_{A,x} + \lambda h  G_{A,y} - \half \lambda  F_{A,xx} 
- \mfrac{3}{2}\lambda  F_{A,yy} + h \lambda  F_{A,xy} . 
\end{align}

Using (\ref{lot78}) and (\ref{78G}) we obtain the system of coupled 
2D NLS equations 
\begin{align}
 2 i \omega F_T  = \; & 3 \beta ( 3 |F|^2 F + 2 |P|^2 F + P^2 F^* )
+ \alpha  (F_{yy} + P_{xy}) , \nn \\
 2 i \omega P_T  = \; & 3 \beta ( 3 |P|^2P + 2 |F|^2 P + F^2 P^*  )
+ \alpha (F_{xy} + P_{xx} ) ,   \qquad \alpha = 
\displaystyle \frac{2 \lambda\gamma}{\gamma+\lambda} . 
\lbl{NLS2-FPT} \end{align}
We are not aware of systems of this form having been derived 
or studied before. 
Clearly there are solutions of the form $P=0$ and $F=F(y,T)$ independent of $x$
satisfying a 1D NLS equation, or $F=0$ and $P=P(x,T)$ independent of $y$ 
and satisfying 1D NLS, which give rise to localised solutions such as  $P=A 
\ee^{- 9 i \beta A^2 T/4\omega} \sech(Ax\sqrt{\beta/2\alpha})$). However, 
it is genuinely two-dimensional solutions that are of greater interest. 

\subsection{\blue{Properties of the coupled NLS system}} \lbl{prop-sec}

\blue{The system of two coupled 2+1d NLS equations (\ref{NLS2-FPT}) 
has the conserved quantity 
\beq
N = I_1 = \int\!\!\int |F|^2+|P|^2 \,\dd x \, \dd y , 
\lbl{I1} \eeq
which corresponds to number, or norm, or charge, this being the 
simplest measure of the magnitude of $F,P$.  There is a second 
conserved quantity, corresponding to energy, given by 
\begin{align}
E = I_2 = \int\!\!\int \, & \alpha \left( 
2 |F_y|^2 + 2 |P_x|^2 + F_x P_y^* + F_x P_y^* 
+ F_y P_x^* + F_y^* P_x \right) \nn \\ & 
- 3 \beta \left( 3 |F|^4 + 3 |P|^4 + 4 |F|^2 |P|^2 
+ F^2 P^{*2} + P^2 F^{*2}  \right) \, \dd x\, \dd y . 
\lbl{I2} \end{align}    }
If we seek the dispersion relation of (\ref{NLS2-FPT}), using 
$(F(x,y,T),P(x,y,T))=(A,B)\ee^{iKx+iLy+i\widetilde\Omega T}$ we find 
two surfaces, which we write as $\widetilde\Omega(K,L)$ as 
\begin{align}
\widetilde\Omega(K,L) = \; & \widetilde\Omega_{\mbox{{\scriptsize flat}}}(K,L) 
= - \frac{9\beta(A^2+B^2)}{2\omega} , \nn \\    
\widetilde\Omega(K,L) = \; & \widetilde\Omega_{\mbox{{\scriptsize opt}}}(K,L) 
= - \frac{9\beta(A^2+B^2)}{2\omega} + \frac{\alpha(K^2+L^2)}{2\omega} . 
\end{align} 
Thus, there is a flat band (where $\widetilde\Omega$ is independent of 
wavenumbers $K,L$) and a wavenumber-dependent band, 
which corresponds to the optical (or middle) band of the
of the original kagome lattice.  
In the case of arbitrarily small waves ($A,B\rightarrow0$) 
$\widetilde\Omega>0$ (assuming $\alpha>0$, $K,L\neq0$). 

We follow the usual method for finding spatially localised 
solutions of a NLS system by writing 
\begin{align}
F(x,y,T) = \; & \ee^{i \widetilde\Omega T} \widetilde F(x,y) , \qquad  
P(x,y,T) = \; \ee^{i\widetilde\Omega T} \widetilde P(x,y) ,  \lbl{FPans}
\end{align}
then (\ref{NLS2-FPT}) are transformed to the coupled PDE system
\begin{align}
- 2 \widetilde\Omega \omega \widetilde F  = \; & 9\beta\widetilde F (\widetilde F^2+\widetilde P^2)
+ \alpha (\widetilde F_{y} + \widetilde P_{x})_y   , \nn \\
 -2 \widetilde\Omega \omega \widetilde P  = \; & 9\beta\widetilde P(\widetilde P^2+\widetilde F^2 )
+ \alpha  (\widetilde F_{y} + \widetilde P_{x} )_x .  \lbl{PDE2} 
\end{align}
In Section \ref{simil-sec} we show how (\ref{PDE2}) 
can be reduced to give other similarity solutions.  

\section{Similarity reductions} \lbl{simil-sec} 
\setcounter{equation}{0} 

\subsection{{\bf Lie symmetry analysis } \lbl{lie-sec}}

We now consider the derivation of similarity reductions associated 
to classical Lie point symmetries \cite{BK89,O93,S89} for the system 
of equations (\ref{PDE2}). In order to do so we require the invariance 
of this system under the one-parameter Lie group of infinitesimal 
transformations in $(x,y,\widetilde F,\widetilde P)$ given by
\beqa
& & x\rightarrow x+\epsilon\, \xi(x,y,\widetilde F,\widetilde P)+O(\epsilon^2), \\
& & y\rightarrow y+\epsilon\, \tau(x,y,\widetilde F,\widetilde P)+O(\epsilon^2), \\
& & \widetilde F\rightarrow \widetilde F+\epsilon\, \phi_1(x,y,\widetilde F,\widetilde P)+O(\epsilon^2),\\
& & \widetilde P\rightarrow \widetilde P+\epsilon\, \phi_2(x,y,\widetilde F,\widetilde P)+O(\epsilon^2),
\eeqa
where $\epsilon$ is the group parameter. The symmetry generator 
associated to the above group of point transformations can be written as
\beq
{\bf v}=\xi(x,y,\widetilde F,\widetilde P) \,\frac{\partial}{\partial x}+\tau(x,y,\widetilde F,\widetilde P)\, 
\frac{\partial}{\partial y}+
       \phi_1(x,y,\widetilde F,\widetilde P)\, \frac{\partial}{\partial \widetilde F}+\phi_2(x,y,\widetilde F,\widetilde P)\, 
       \frac{\partial}{\partial \widetilde P}.
\eeq
The invariance condition leads to an overdetermined system of 
linear differential equations (the determining equations) for the 
infinitesimals $\xi, \tau$, $\phi_1$ and $\phi_2$. Once the infinitesimals 
have been obtained,  the similarity variables are found by solving 
the associated characteristic equations
\beq
\frac{dx}{\xi(x,y,\widetilde F,\widetilde P)}=\frac{dy}{\tau(x,y,\widetilde F,\widetilde P)}=
\frac{d\widetilde F}{\phi_1(x,y,\widetilde F,\widetilde P)}=\frac{d\widetilde P}{\phi_2(x,y,\widetilde F,\widetilde P)}. 
\lbl{char} 
\eeq
The infinitesimals $\xi, \tau$, $\phi_1$ and $\phi_2$ associated to the classical 
Lie symmetries of the system of equations (\ref{PDE2}) are found to be
\beqa
\xi & =& -c_1y+c_3,\lbl{inf1}\\
\tau & = & c_1x+c_2,\lbl{inf2}\\
\phi_1 & = & c_1 \widetilde P,\lbl{inf3}\\
\phi_2 & = & -c_1 \widetilde F \lbl{inf4}
\eeqa
where  $c_1, c_2$ and $c_3$  are arbitrary constants. Depending on the choice 
of the constant $c_1$, we obtain two different similarity reductions. 

\subsection{{\bf First similarity reduction } \lbl{trav-wave}}

First of all, for the choice $c_1=0$, we may set without lost of generality 
$c_2=1$ and relabel $c_3=c$. We then easily obtain the \blue{plane} wave reduction 
\begin{align}
\widetilde P(x,y)=u(z) , \qquad \widetilde F(x,y) = v(z) , \qquad  z=x-cy. 
\lbl{PFuv-trav} \end{align}
The associated system of ODEs is
\begin{align}
0 = &  c^2v''-cu''+\frac{9\beta}{\alpha} v(v^2+u^2)+\frac{2\widetilde\Omega \omega}{\alpha}v, \\
0 = & cv''-u''-\frac{9\beta}{\alpha} u(u^2+v^2)-\frac{2\widetilde\Omega \omega}{\alpha}u,  
\end{align}
which implies 
\beq
(v+cu)\left[\frac{9\beta}{\alpha} (v^2+u^2)+\frac{2\widetilde\Omega \omega}{\alpha}\right]=0
\eeq
For $v=-cu$ both equations reduce to an ODE which can be integrated to give
\beq
u'^2 + \frac{9\beta}{2\alpha}u^ 4 - 
\frac{2\widetilde\Omega \omega}{\alpha (c^2+1)} u^2 + C = 0 ,
\eeq
where $C$ is an arbitrary constant of integration, and which can be solved 
in terms of elliptic or hyperbolic functions. The other possibility simply yields 
$u=cv+\gamma z+\delta$ together with $9\beta (v^2+u^2) + 
2\widetilde\Omega \omega=0$, where again $\gamma$ and $\delta$ 
are two arbitrary constants of integration.

We do not pursue this reduction further here, since these are plane waves 
and can only be localised in one direction, instead we concentrate on 
the choice $c_1\neq0$, which provides localized solutions in both directions ($x$ and $y$). 
This reduction is considered in the next subsection.

\subsection{{\bf Second similarity reduction } \lbl{sol-sec}}

In the case $c_1\neq0$ we can take $c_2=c_3=0$ without lost of generality. 
The characteristic equations are 
\beq
\frac{dx}{-y}=\frac{dy}{x}=\frac{d\widetilde F}{\widetilde P}=\frac{d\widetilde P}{-\widetilde F},
\eeq
which implies $r^2=x^2+y^2$ and $R^2=\widetilde F^2+\widetilde P^2$, where $r$ 
and $R(r)$ are constants of integration. We also have
\beq
\frac{dy}{x}=\frac{d\widetilde F}{\widetilde P},
\eeq
which in terms of $r$ and $R$ is written as
\beq
\frac{dy}{\sqrt{r^2-y^2}}=\frac{d\widetilde F}{\sqrt{R^2-\widetilde F^2}}.
\eeq
Integrating this gives
\beq
\widetilde F=\frac{R(r)}{r}(y\cos (M(r))+x\sin (M(r)),
\eeq
with $M(r)$ being the third constant of integration. Now, using 
the fact that $R^2=\widetilde F^2+\widetilde P^2$, we get
\beq
\widetilde P=\frac{R(r)}{r}(x\cos (M(r))-y\sin (M(r)).
\eeq
We thus obtain a similarity reduction of (\ref{PDE2}) 
of the form 
\begin{align}
\widetilde P(x,y)= x u(r) + y v(r) , \qquad 
\widetilde F(x,y) = y u(r) - x v(r) , \qquad r^2=x^2+y^2 , 
\lbl{PFuv-simil}
\end{align}
where we have defined new variables $u(r),v(r)$ by  
$u(r)=R(r)\cos (M(r))/r$ and $v(r)=-R(r)\sin (M(r))/r$. Substituting 
this similarity reduction into the system (\ref{PDE2}), we obtain
\begin{align}
0 = &  u''+\frac{3}{r}u'+\left(u-\frac{xv}{y}\right)\left[ 
\frac{9\beta}{\alpha}r^2(u^2+v^2)+\frac{2\widetilde\Omega\omega}{\alpha}\right] , \\
0 = & u''+\frac{3}{r}u'+\left(u+\frac{yv}{x}\right)\left[
\frac{9\beta}{\alpha}r^2(u^2+v^2)+\frac{2\widetilde\Omega\omega}{\alpha}\right].  
\end{align}
Adding and subtracting these last two equations, we get the pair of equations
\begin{align}
0 = & \; \left[ 2 \omega \widetilde\Omega  + 9 \beta  r^2 (u^2+v^2)\right]\left[ 
u+\frac{v(y^2-x^2)}{2xy}\right] 
+\alpha \left(u''+\frac{3}{r} u' \right) , \\
0 = & v \left[  2 \omega \widetilde\Omega  + 9 \beta r^2 (u^2+v^2) \right] .  
\end{align}
Solving the latter equation first, we either have $v=0$ or 
$u^2 + v^2 = - 2\omega \widetilde\Omega / (9 \beta r^2)$. 
With the latter solution, the equation for $u$ is $u'' + (3/r) u'=0$ 
which implies $u = A + B r^{-2}$, which is in general 
both singular at the origin, and non-zero as $r\rightarrow\infty$. 
In addition, this does not lead to a well-defined solution for $v(r)$ since 
\beq
v^2 = - \frac{2 \omega \widetilde\Omega}{9 \beta r^2} - A^2 
- \frac{2AB}{r^2} - \frac{B^2}{r^4} ,  
\eeq
which is negative at large and small values of $r$.  

Using the former and simpler solution $v=0$ leads to a more 
complicated equation for $u$, namely 
\beq
0 = \alpha \left( u'' +  \frac{3}{r} u' \right) + 9 \beta r^2 u^3 
+ 2 \omega\widetilde\Omega u ,  \lbl{u-ode} 
\eeq
which in the general case does not appear to have explicit solutions 
in terms of elementary functions, 
but some of whose properties are available, and in the next subsection 
we give a method for constructing approximate solutions. 
If we consider the form of $u(r)$ at large values of $r$, 
we expect $u(r) \rightarrow 0$.  If we were to assume that the decay 
is exponential, with $u \sim \ee^{-\lambda r}$ then  we find 
$\lambda^2 = - 2 \omega\widetilde\Omega/\alpha$, thus we might expect 
$\widetilde\Omega<0$ in (\ref{FPans}), given that $\omega>0$ and $\alpha>0$. 
However, in general, the decay may only be algebraic, 
in which case, no such simple inequality holds. 
\blue{ In terms of the original variables $q^A,q^B,q^C$, 
inverting the transformations (\ref{PFuv-simil}), (\ref{lot78}), 
and (\ref{ansatz})--(\ref{xytauT}) leads to 
\begin{align}
q^A_{m,n} = \; & 4 \sqrt{3} \ep^2   n 
\cos( (\omega-\ep^2\widetilde\Omega) t ) \, u( \ep^2(m^2+3n^2) ) , \nn \\
q^B_{m,n} = \; & -2 \ep^2 (n+m\sqrt{3}) 
\cos( (\omega-\ep^2\widetilde\Omega) t ) \, u( \ep^2(m^2+3n^2) ) , \nn \\
q^C_{m,n} = \; & -2 \ep^2 (n-m\sqrt{3}) 
\cos( (\omega-\ep^2\widetilde\Omega) t ) \, u( \ep^2(m^2+3n^2) ) . 
\lbl{holistic} \end{align} } 
If we consider the form of $u(r)$ at large values of $r$, 
we expect $u(r) \rightarrow 0$. 

\subsection{{\bf Weakly nonlinear approximate solution} \lbl{weak-nl-sec}}

In the limit of small nonlinearity ($\beta\rightarrow0$), equation 
(\ref{u-ode}) 
has an explicit solution of the form $u(r) = (c/r) J_1 ( r \sqrt{2\omega 
\widetilde\Omega/\alpha})$, with $c$ being an arbitrary constant,  a solution 
which requires $\widetilde\Omega>0$. 
Following the theory developed by Benjamin \cite{tbb67} and Whitham 
\cite{whitham}, for small positive $\beta$ ($0<\beta\ll1$), we generalise 
$c$ to \blue{a} function which depends on a new \blue{``slow''}  
variable $\rho=\beta r$ 
\beq
u(r,\rho) = \frac{ c(\rho)}{r} J_1 \left( r \sqrt{ 
\frac{2\omega \widetilde\Omega}{\alpha}} \right) + \beta q(r) .  \lbl{tbb} 
\eeq
We determine 
the form of the arbitrary `function' $c(\rho)$ by requiring a 
secularity condition on the problem for the first correction term, $q(r)$, 
Substituting (\ref{tbb}) into (\ref{u-ode}), and expanding, we obtain 
\begin{align}
\alpha \left( \frac{\dd^2 q}{\dd q^2} + \frac{3}{r} \frac{\dd q}{\dd r} 
\right) + 2 \omega \widetilde\Omega q = \; & R(r) , \lbl{Rode}
\end{align}
where 
\begin{align} 
R(r) := \; & 
- 9 \beta r^2 u^3 - \alpha  \beta \left( 2 \frac{\dd c}{\dd \rho} \left( 
\frac{\sqrt{2\omega\widetilde\Omega}}{r\sqrt{\alpha}} J_0\left( 
r\sqrt{\frac{2\omega\widetilde\Omega}{\alpha}} \right) - 
\frac{2}{r^2} J_1\left(r\sqrt{\frac{2\omega\widetilde\Omega}{\alpha}}\right) \right) + 
\frac{3}{r^2} \frac{\dd c}{\dd \rho}  J_1 \left( r 
\sqrt{\frac{2\omega\widetilde\Omega}{\alpha}} \right) \right) . 
\end{align}
Here, we have omitted the $\mathcal{O}(\beta^2)$ term as this is 
a higher order correction term. 

\begin{figure}[hbtb] 
\includegraphics[scale=0.4]{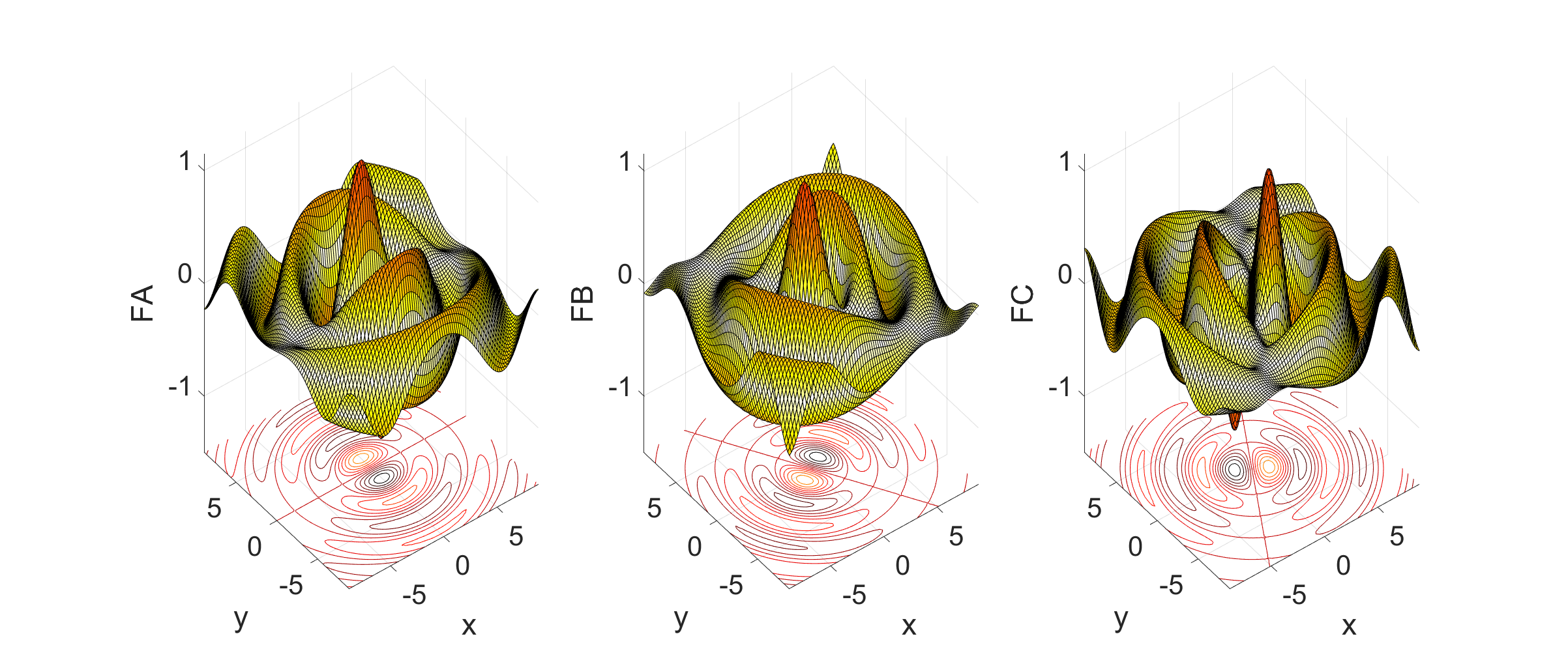} 
\caption{ Illustration of the vector field solutions 
for $F_A,F_B,F_C$ generated by  (\ref{tbb}) 
and $(\widetilde P(x,y),\widetilde F(x,y))
 = (x u(r),y u(r))$ with $r^2=x^2+y^2$  in the case $v=0$, 
 for the parameter values $\alpha=1$, $\omega\widetilde\Omega=1$,  
 $\beta=0.1$,  (in colour in on-line version).   \lbl{vec-fig} }
\end{figure}

A Fourier-Bessel series has the form $f(x) \sim \sum_{n=1}^\infty a_n 
J_\nu ( \lambda_n x)$ and the corresponding orthogonality condition is with 
respect to the inner product $\langle f(r),g(r) \rangle=\int_0^\infty r f(r) g(r)\dd r$.  
Thus, in order for (\ref{Rode}) to have a solution, we impose the condition 
$\int_0^\infty r J_1(r\sqrt{2\omega\widetilde\Omega}) R(r) \dd r=0 $, 
which implies 
\beq  
0 = \left( 1 +\frac{\sqrt{\alpha}}{2\sqrt{2\omega\widetilde\Omega}} \right) 
\frac{\dd c}{\dd \rho} +\frac{\sqrt{2\alpha}}{4\pi\sqrt{\widetilde\Omega\omega}} M c^3 ,  
\lbl{dcdro} \eeq
where $M= G([[-\half,\half],[\half,\mfrac{3}{2}]],[[1,0],[0,-1]],1) \approx 1.258$ 
is a constant found from the MeijerG function \cite{maple,olver}. 
Solving equation (\ref{dcdro}) subject to the 
initial condition $c(0)=1$ leads to a solution $c(\rho)$, 
and hence $u(r)$ of the form 
\beq
c(\rho) = \frac{1}{\sqrt{1 + K \rho }}  , \qquad 
K = \frac{2 \sqrt{2\alpha}}{\pi ( \sqrt{2\alpha} 
+ 4 \sqrt{2\omega\widetilde\Omega} ) } , 
\qquad   u(r) = \frac{1}{ r \sqrt{1+ \beta K r}} J_1\left( r 
\sqrt{\frac{2\omega\widetilde\Omega}{\alpha}} \right) . 
\lbl{cKu} \eeq
The corresponding solutions for $F_A,F_B,F_C$ generated from $(\widetilde P,\widetilde F) 
= (x u(r),y u(r))$ are plotted in Figure \ref{vec-fig}. Whilst they have similar 
shapes, they are not cylindrically symmetric, but each is a rotation of another 
by 120$^\circ$.  Whilst these appear highly oscillatory and only decaying 
slowly in space, when superimposed, and plotted as intensity 
($|F|^2$ instead of $F$), as shown in Figure \ref{vec-fig2},  
they more closely resemble vortex solitons observed in other systems. 
For example, the numerical simulations of 2D NLS system of Ablowitz 
\etal\ \cite{mja} which has includes a Penrose lattice potential, 
and the 2D NLS model of Zeng and Malomed  \cite{bam} which 
has a lattice structure induced by periodic modulation 
of both the linear and nonlinear potentials.  

\begin{figure}[hbtb] 
\hspace*{-5mm}
\includegraphics[scale=0.6]{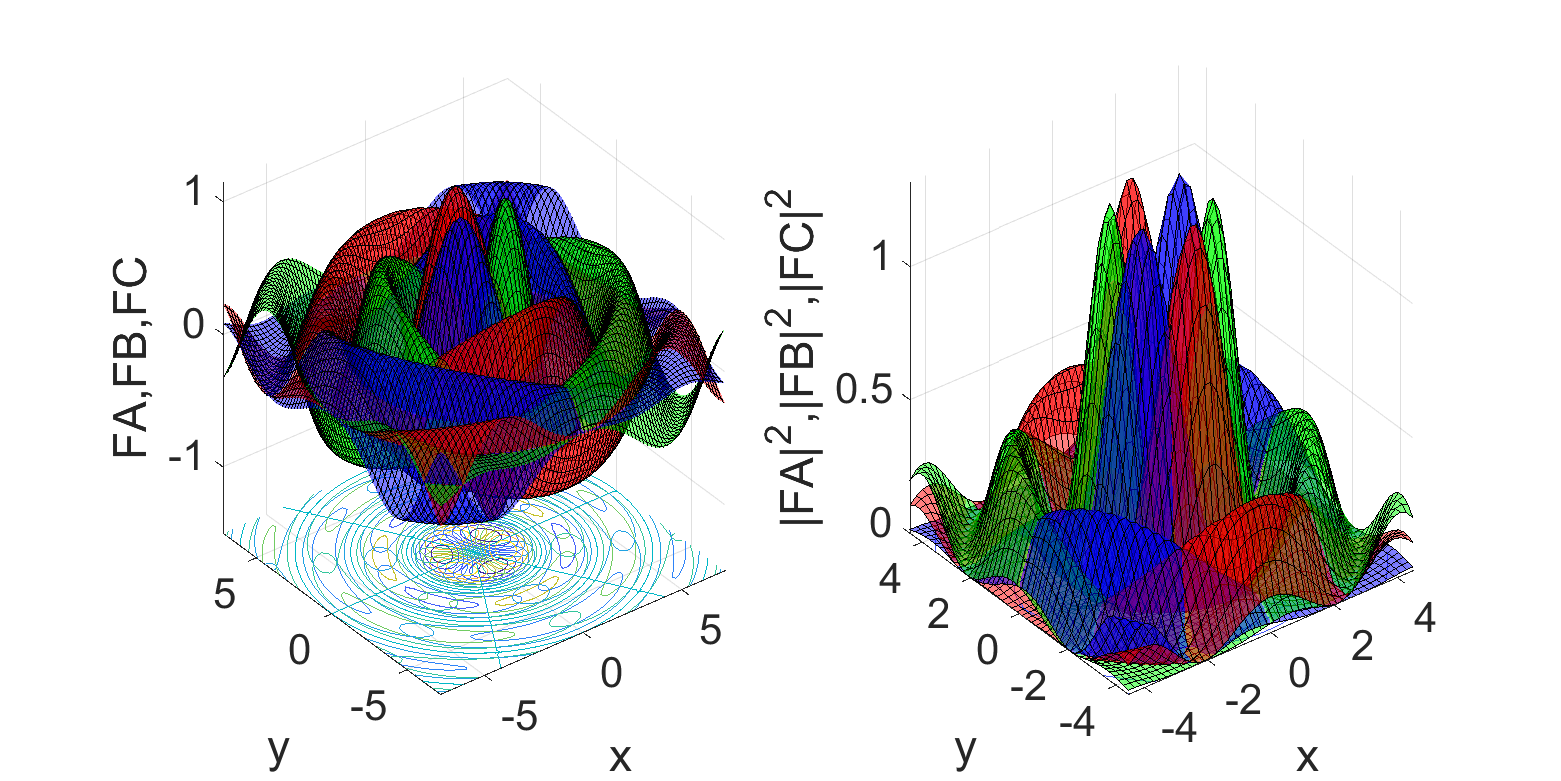} 
\caption{ Illustration of the vector field solution 
for $F_A,F_B,F_C$ generated by (\ref{tbb}) and 
$(\widetilde P(x,y),\widetilde F(x,y))
 = (x u(r),y u(r))$ with $r^2=x^2+y^2$  in the case $v=0$, 
 for the parameter values $\alpha=1$, $\omega\Omega=1$, $\beta=0.1$.  
(In colour on-line, the three functions $F_A,F_B,F_C$ being 
plotted in red, green, blue.)
\lbl{vec-fig2} }
\end{figure}

\blue{We illustrate the mode, by presenting the results 
of simulations in Figure \ref{num-fig}. These show the case 
of $\Omega=2$, $\lambda=1/2=\gamma$, $k=0=l$, $\beta=-0.1$, 
on a lattice of size $160\times 80$.  
The kagome lattice structure of the equilibrium node positions 
can be seen as black circles, with red lines indicating 
displacements at the end of a simulation. }
\blue{The energy in each unit cell of the lattice is defined by 
\begin{align}
e_{m,n} = \; & \frac12 \left( \!\frac{\dd q^A_{m,n}}{\dd t} \right)^2 + 
\frac12 \left( \!\frac{\dd q^B_{m,n}}{\dd t} \right)^2  + 
\frac12 \left( \!\frac{\dd q^C_{m,n}}{\dd t} \right)^2  + 
\frac{ \Omega^2}{2} \left( (q_{m,n}^{A})^2+(q_{m,n}^{B})^2+(q_{m,n}^{C})^2 
\right) \nn \\ &   + \frac{\beta}{4} \left( 
(q_{m,n}^{A})^4+(q_{m,n}^{B})^4+ (q_{m,n}^{C})^4 \right) 
+ \frac{\gamma}{2} \left[  (q_{m,n}^A - q_{m,n}^B )^2 
+ ( q_{m,n}^B - q_{m,n}^C )^2 + ( q_{m,n}^C - q_{m,n}^A )^2 \right] 
\nn \\ &   
+ \frac{\lambda}{2} ( q_{m,n}^A - q_{m+1,n+1}^B )^2
+ \frac{\lambda}{2} ( q_{m,n}^A - q_{m-1,n+1}^C )^2
+ \frac{\lambda}{2} ( q_{m,n}^B - q_{m-2,n}^C )^2 , 
\lbl{en} \end{align}
which, when summed over the lattice, gives the Hamiltonian, 
(\ref{ham}). 
On the right of Figure \ref{num-fig} we present 3D plots 
of the energy $e_{m,n}(t)>0$ both at the start (centre-right panel) 
and end (upper right panel) of the simulation.  The energy distribution 
is not identical, but is very similar, suggesting the mode is robust. 
\purple{The time period for the flat band is $t_{\mbox{{\scriptsize period}}} = 
2\pi/\omega=2.3748$, so the final time of $t_{\mbox{\scriptsize final}} 
= 200$ corresponds to over 80 oscillations of the carrier wave. 
The lower panel of Figure \ref{num-fig} shows a cross-section 
of the lattice through $m=0$, where $q^K_{0,n}$ is plotted against 
$n$ for $K\in\{A,B,C\}$ at times $t=0$ -- where $q^K_{m,n}$ 
are given by the asymptotic approximations  (\ref{cKu}), (\ref{PFuv-simil}), 
(\ref{lot78}), (\ref{ansatz})--(\ref{xytauT}).  
The equations of motion have then been numerically 
integrated forward in time till $t_1 =146.725$ - approximately 62 
oscillations of the carrier wave, 
and the displacements $q^K_{0,n}(t_1)$ plotted 
on the same axes.  A small difference in the displacements
$q^A$ can be seen at small $n$, but these differences decrease 
at larger $n$, and are also very small for $q^B, q^C$.  } 
This simulation is on a relatively small grid, (160$\times$80), so the 
breather mode has not fully decayed at the edges of the grid.  
Since the predicted wave only decays algebraically, with 
$u\sim r^{-2}$ for $r\gg1$, an extremely large lattice would be 
needed to accurately simulate the solitary wave. 
Furthermore, only a weakly nonlinear approximation 
to the leading order term has been used as initial conditions, 
and so some initial transient adjustment of wave is to be expected. }

\begin{figure}[hbt] 
\includegraphics[scale=0.4]{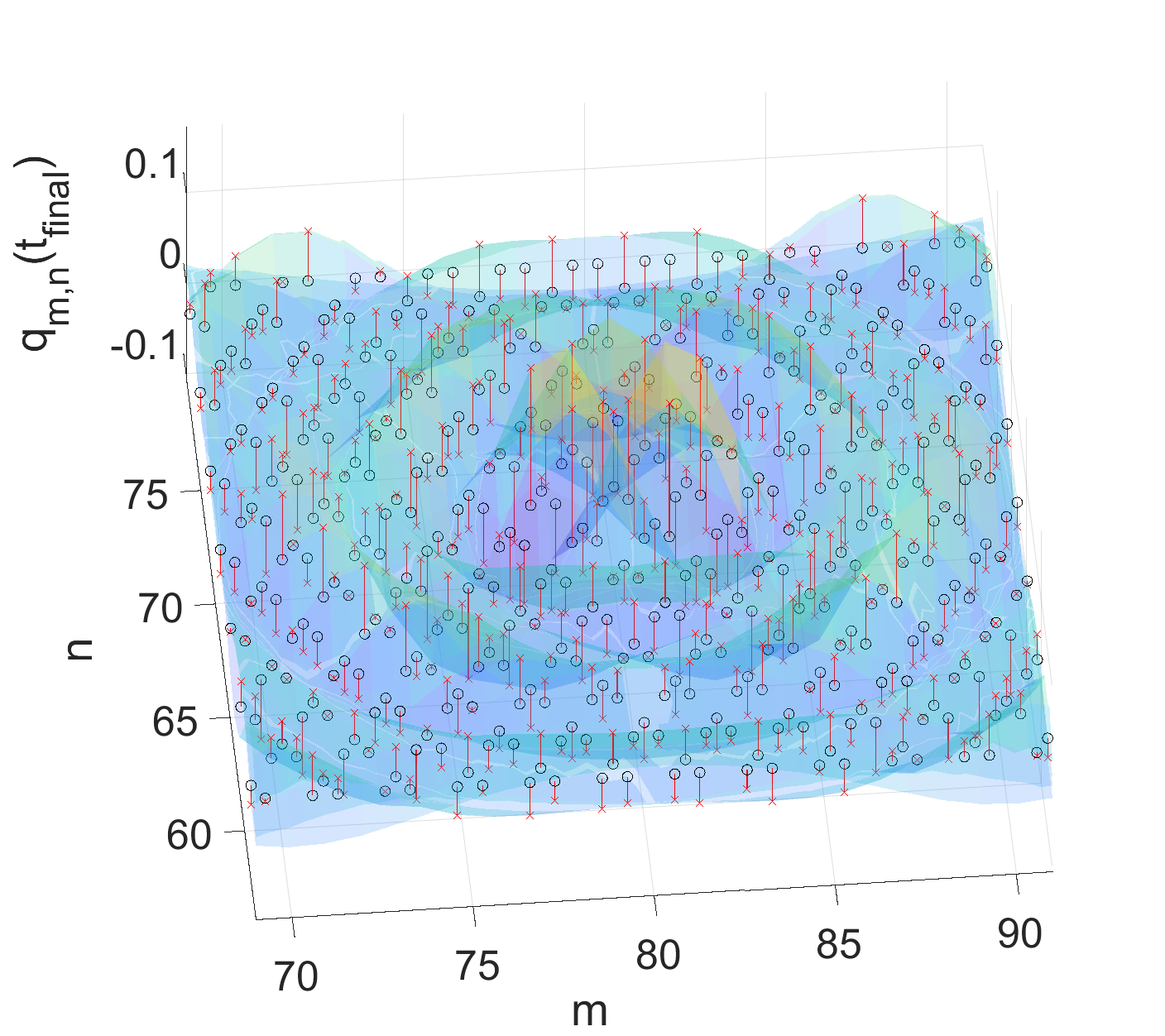} \hspace*{-6mm} 
\raisebox{44mm}{\includegraphics[scale=0.42]{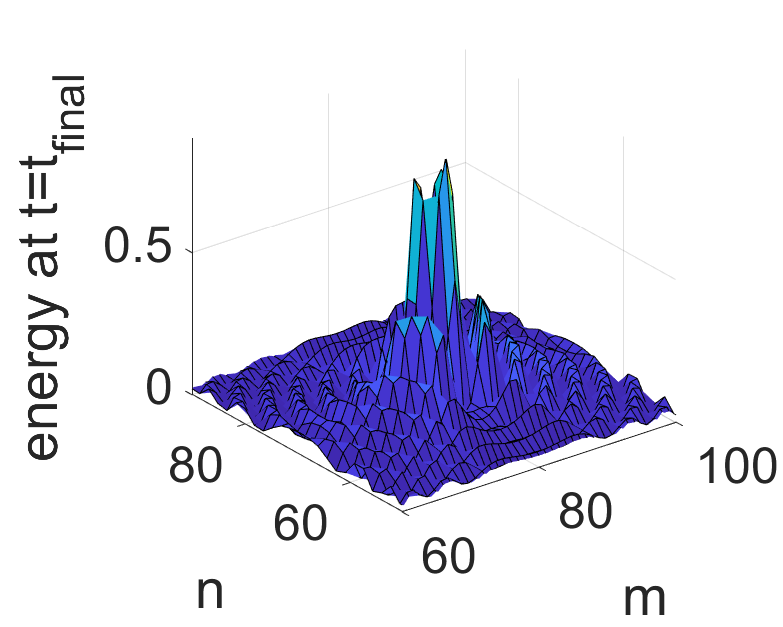}} 
\hspace*{-60mm}
\raisebox{-2mm}{\includegraphics[scale=0.42]{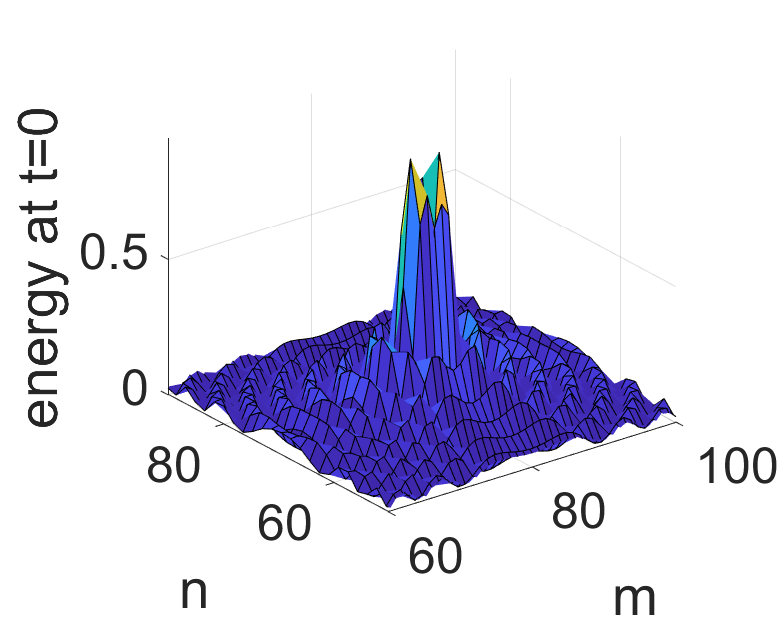}} 
\hspace*{-5cm} 
\newline
\includegraphics[scale=0.5]{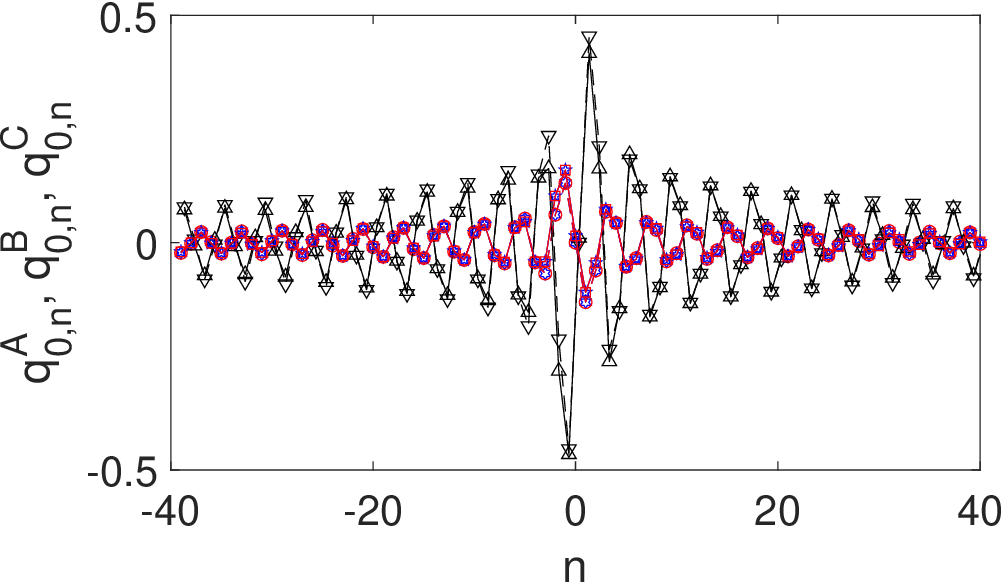}
\caption{\blue{Results of numerical simulations of breather mode 
with initial conditions given by (\ref{cKu}), (\ref{PFuv-simil}), 
(\ref{lot78}), (\ref{ansatz})--(\ref{xytauT}), then numerically integrated 
forward in time using (\ref{qAddot})--(\ref{qCddot}). 
Upper left:  plots of $q^K_{m,n}(t_{\mbox{\scriptsize $m,n$}})$, 
($K=A,B,C$), circles show equilibrium positions of nodes 
($q_{m,n}^K=0$), red crosses show displacements at 
$t=t_{\mbox{{\scriptsize final}}}$, with lines from $q=0$.  
The semi-transparent surfaces show the smoothed envelope 
for each of the $q^A$, $q^B$, $q^C$ amplitudes. 
Centre right: the energy distribution (\ref{en})  at $t=0$ -- as 
given by the asymptotic approximation (\ref{holistic}).  
Upper right: the energy distribution after numerically integrating 
the ODEs for $q^K(t)$ (\ref{qAddot})--(\ref{qCddot}) forward in 
time to $t=t_{\mbox{{\scriptsize final}}}=200$ using the Verlet algorithm. 
\purple{Lower panel: plot of lattice displacements, with the initial conditions 
given by the asymptotic approximation, denoted by dashed lines 
and black inverted triangles for $q^A_{0,n}(0)$, blue hexagrams for
$q^B_{0,n}(0)$, red circles for $q^C_{0,n}(0)$;  we also plot 
the displacements near the end of the simulation, at $t_1=146.725$, 
where $q^A_{0,n}(t_1)$ is plotted in black triangles, 
$q^B_{0,n}(t_1)$ is plotted in blue pentagrams, 
$q^C_{0,n}(t_1)$ is plotted in red squares. } 
In colour in on-line version. }   \lbl{num-fig}}
\end{figure}

\blue{The issue of stability, remains open. 
We compute the norm $N(\widetilde\Omega)$
from (\ref{I1}) and determine its dependence on the 
frequency $\widetilde\Omega$, making use of 
$F = \ee^{i\widetilde\Omega T} y u(r)$, 
$P = \ee^{i\widetilde\Omega T} x u(r)$, $r^2=x^2+y^2$, 
together with (\ref{cKu}) and (\ref{PFuv-simil}).  
We find that $N(\widetilde\Omega)$ is a monotonically decreasing 
function of frequency, $\widetilde\Omega$.  If the stability results 
of Vakhitov-Kolokolov \cite{kolo,vk} apply to the coupled NLS system 
(\ref{NLS2-FPT}), then this results suggests the solution is not stable. 
However, the stability of such a mode in the original lattice system 
(\ref{qAddot})--(\ref{qCddot}) is a different problem, which remains open.  }

\subsection{{\bf Other coupled NLS systems} \lbl{other-sec}}

In this section, we consider extensions to the system of PDEs 
(\ref{NLS2-FPT}) in order to find the most general set of two 
coupled NLS equations which allows a reduction to ODEs 
using the similarity substitution (\ref{PFuv-simil}).  
We include other second spatial derivative terms and 
other cubic nonlinear combinations of $F,P$. 
The most general form of two coupled NLS equations that 
we have found is 
\begin{align}
2i\omega F_T = \; & (\beta+\eta) |F|^2 F + \beta |P|^2 F + \eta P^2 F^*
+ c_1 (F_y+P_x)_y + c_2 (F_y+P_x)_x \nn \\ & 
+ c_3 (P_y-F_x)_y + c_4 (P_y-F_x )_x + c_5( F_{xx}+F_{yy}) ,  \nn \\ 
2i\omega P_T = \; & (\beta+\eta) |P|^2 P + \beta |F|^2 P + \eta F^2 P^* 
+ c_1 (F_y+P_x)_x  - c_2 (F_y+P_x)_y \nn \\ & 
+ c_3 (P_y-F_x)_x - c_4 (P_y-F_x)_y + c_5 (P_{xx}+P_{yy} ) . 
\end{align}
The resulting reduction (\ref{FPans}) with (\ref{PFuv-simil}) removes the 
time-dependence and leads to the PDEs  
\begin{align}
0 = \; & 2\omega \widetilde\Omega \widetilde F + (\beta+\eta) (\widetilde F^2+\widetilde P^2) \widetilde F 
+ c_1 ( \widetilde F_y+\widetilde P_x )_y + c_2 (\widetilde F_y+\widetilde P_x)_x \nn \\ & 
+ c_3 (\widetilde P_y-\widetilde F_x)_y + c_4(P_y-F_x)_x + c_5 (\widetilde F_{xx}+\widetilde F_{yy})  , 
\nn \\ 
0 = \;  & 2\omega\widetilde\Omega \widetilde P + (\beta+\eta)(\widetilde F^2+\widetilde P^2) \widetilde P 
+ c_1 ( \widetilde F_y+\widetilde P_x)_x - c_2 (\widetilde F_y+\widetilde P_x)_y \nn \\ & 
+ c_3 (\widetilde P_y-\widetilde F_x)_x - c_4 (P_y-F_x)_y + c_5 (\widetilde P_{xx} + \widetilde P_{yy}) , 
\end{align}
and using (\ref{PFuv-simil}), 
\begin{align}
0 = \; & (c_1+c_5) (u'' + \mfrac{3}{r} u' ) + c_3 (v'' + \mfrac{3}{r} v') 
+ 2\omega\widetilde\Omega u + (\beta+\eta) r^2 (u^2+v^2) u , \nn \\ 
0 = \;  & (c_5-c_4) (v'' + \mfrac{3}{r} v') - c_2 ( u'' + \mfrac{3}{r} u') 
+ 2\omega\widetilde\Omega v + (\beta+\eta) r^2 (u^2+v^2) v .  \lbl{omit}
\end{align}
which can be written in matrix form as 
\beq   {\bf M}  
\begin{pmatrix} u'' + \mfrac{3}{r} u' \\ v'' + \mfrac{3}{r} v' \end{pmatrix} := 
\begin{pmatrix} c_1+c_5 & c_3 \\ -c_2 & c_5-c_4 \end{pmatrix} 
\begin{pmatrix} u'' + \mfrac{3}{r} u' \\ v'' + \mfrac{3}{r} v' \end{pmatrix} = 
- [ 2\omega\widetilde\Omega + (\beta+\gamma)r^2 (u^2+v^2) ]  
\begin{pmatrix} u \\ v \end{pmatrix} .  \lbl{Muv}
\eeq
The example described in sections \ref{sol-sec}--\ref{weak-nl-sec}) 
corresponds to the case $c_2=c_3=c_4=c_5=0$, 
which makes the matrix in this equation singular, and simplifies 
the solution process allowing us to find the solutions with 
$v=0$ and $u=A+B/r^2$ and (\ref{u-ode}).  

The more general system (\ref{Muv}) can be reduced to a single 
equation by considering the eigenvalues ($\Lambda$) and 
eigenvectors of ${\bf M}$.  Writing the eigenvectors as 
${\bf M}(1,q)^T = \Lambda (1,q)^T$, we have 
\beq
\Lambda = \half \left( 2 c_5 + c_1 - c_4 \pm \sqrt{ 
(c_4+c_1)^2 - 4 c_2 c_3 }\, \right) ,   \qquad 
q = - \frac{1}{2 c_3} \left( c_1 + c_4 \pm \sqrt{ 
(c_1+c_4)^2 - 4 c_2 c_3 } \,\right) , 
\eeq
\beq
0 = u'' + \frac{3}{r} u' + \frac{ 2 \omega\widetilde\Omega + 
(\beta+\eta)(1+q^2) r^2 u^2  }{ (c_1 + c_5 + q c_3) } \, u . 
\eeq
Weakly nonlinear solutions of this can be found using the 
same methods as explained in Section \ref{weak-nl-sec}. 

\section{Conclusions} \lbl{conc-sec}
\setcounter{equation}{0}

We have analysed the kagome lattice with Klein-Gordon interactions, 
that is, linear nearest neighbour interactions, and a nonlinear on-site 
potential.  The simple case of only a scalar unknown quantity at each 
node has been considered here, in future work, we plan to 
investigate the effects of this geometry in a mechanical system 
in which two displacements  \blue{exist at each node}. 
\blue{We assume small amplitude oscillations, and use 
an asymptotic expansion, together with the Fredholm 
alternative to find solutions of systems of equations; 
our results are valid in the weakly nonlinear limit. }
We find the dispersion relation for this system has three surfaces: 
a flat band and two sets of wave-number dependent modes. 
The surfaces meet at Dirac points (that is, in a cone-like fashion), 
as well as a tangential meeting.  We have analysed these special 
cases in detail using asymptotic techniques to reduce the equations 
of motion to (coupled) nonlinear Schr\"{o}dinger systems. 
In simple cases, this gives a single equation in 2+1 dimensions, 
which has the classic Townes soliton solution \cite{Townes}; 
in more complicated cases, we have found similarity solutions 
of the resulting system of NLS equations. 

We see different behaviour at points where dispersion 
surfaces meet tangentially from those where the meeting 
is at single isolated Dirac points. 
At the tangential meeting, we have obtained a novel 
coupled system of NLS equations in 2+1 dimensions (\ref{NLS2-FPT}); 
\blue{we have shown that this system has conserved quantities 
corresponding to `mass' (or charge, corresponding to the $L_2$ norm) 
and energy, given by (\ref{I1})--(\ref{I2}). }
Using similarity 
analysis we have obtained a type of vortex solitary wave solution. 
\blue{We have simulated this wave numerically, showing that 
it is robust, in that it persists for a significant time, 
though extensive numerical simulations lie beyond the scope of this paper
and questions of stability remain open. } 
\blue{Since we expect other, more general, lattices 
to also have flat bands and regions in their dispersion relations 
where surfaces meet tangentially, we expect similar phenomena 
to occur more widely, for example see the 
recent work of Hofstrand \cite{hof26} on a similar lattice,}  
and the super-kagome lattices considered by Kerner \etal\ 
\cite{KTW} which has six nodes in each cell 
of the lattice (rather than three), so this super-kagome lattice would be 
more complicated to analyse than the simple kagome lattice considered 
here due to increased numbers of surfaces in the dispersion relation. 
Whilst similar techniques could be used as here, analytical progress 
would be more difficult due to the larger matrices involved. The system 
will still have stationary modes and Dirac points as seen here, there 
is the possibility for more exotic modes if three surfaces of the 
dispersion relation were to coincide, such systems are left as future work. 
In future work we aim to address more 
complicated kagome lattices, such as the FPUT case with 
nonlinear interactions and no on-site potential 
and mechanical cases where nodes are displaced 
in both directions within the lattice. 

\subsection*{{\bf Acknowledgements}}

We are grateful to the Agencia Estatal de Investigacion (España) for 
supporting Project PID 2020 - 115273 GB - I00 funded by 
MCIN/AEI/ 10.13039 / 501100011033, and also for Grant 
RED2022-134301-T 
funded by MCIN/AEI/10.13039/501100011033. We also gratefully 
acknowledge financial support from the Universidad Rey Juan Carlos 
 under Project 2025/SOLCON-160677 and 
as members of the Grupo de investigación de alto rendimiento DELFO.
We are grateful for funding the publication of this article under the 
transformative agreement between the University of Nottingham 
and Elsevier.  \blue{We are grateful to the referees for suggestions 
on improving the manuscript. }

\subsection*{{\bf Ethics, Integrity, and Conflict of Interests statements}}

The authors know of no conflicts of interest in the undertaking or publication 
of this work.  No use of generative AI has been made in the writing of the 
manuscript.  The presented work does not rely on any other data. 


\end{document}